\documentclass[journal]{IEEEtran}

\usepackage{amssymb}
\usepackage{amsmath}

\usepackage{xcolor}
\usepackage{array}

\usepackage{cite}
\usepackage{url}
\usepackage{hyperref}
\usepackage[normalem]{ulem}
\hypersetup{
	colorlinks=true, 
	linkcolor=blue!70!black, 
	citecolor=red!70!black, 
	filecolor=blue!70!black, 
	urlcolor=blue!70!black, 
}

\usepackage{tikz}
\usepackage{tikz-3dplot} 
\usepackage{animate}
\usepackage{ifthen}
\usepackage{multido}
\usepackage{pgfplots}
\usepackage{calc}

\usepackage{subfigure}
\usepackage{enumitem}
\usepackage{esint} 
\usepackage{booktabs} 
\usepackage{multirow}

\newcommand{\J}{\mathrm{j}}                 
\newcommand{\Quot}[1]{``{#1}"}              
\newcommand{\degree}{\ensuremath{^\circ}}

\newcommand{\RE}{\mathrm{Re}}               
\newcommand{\IM}{\mathrm{Im}}               

\providecommand{\D}{\,\mathrm{d}}           
\providecommand{\V}[1]{\boldsymbol{#1}}     
\providecommand{\M}[1]{\mathbf{#1}}         
\newcommand{\UV}[1]{\hat{#1}}               
\newcommand{\ABS}[1]{\left| #1 \right|}     
\newcommand{\OP}[1]{{\mathcal{#1}}}         

\providecommand{\herm}{\mathrm{H}} 
\providecommand{\trans}{\mathrm{T}}

\providecommand{\JBSL}[2]{\mathrm{j}_{#1} \left( #2 \right)}  
\providecommand{\YBSL}[2]{\mathrm{y}_{#1} \left( #2 \right)}

\providecommand{\srcRegion}{\varOmega} 
\providecommand{\basfcn}{\, \V{\psi}}       
\providecommand*{\EPS}{\varepsilon} 
\providecommand*{\MUE}{\mu} 
\providecommand{\ZVAC}{Z_0}  
\providecommand{\Ivec}{\M{I}}

\providecommand{\CMC}{C_{pq} \left(ka\right)}
\providecommand{\SPC}{\gamma_p \left(ka\right)}

\newcommand{\ie}{\textit{i.e.}{}}

\newcommand{\cf}{\textit{cf.}{}}

\newcommand{\BE}{\begin{equation}}
\newcommand{\EE}{\end{equation}}
\newcommand{\BEn}{\begin{equation*}}
\newcommand{\EEn}{\end{equation*}}
\newcommand{\BF}{\begin{figure}\centering}
\newcommand{\EF}{\end{figure}}
\newcommand{\BT}{\begin{table}\centering}
\newcommand{\ET}{\end{table}}

\newcommand\figwidth{9}
\pgfplotsset{compat=1.14} 

\begin{document}
\title{Validating the Characteristic Modes Solvers}
\author{Miloslav~Capek,~\IEEEmembership{Member,~IEEE,}
		Vit~Losenicky,
        Lukas~Jelinek,
        and~Mats~Gustafsson,~\IEEEmembership{Member,~IEEE}
\thanks{Manuscript received January XX, revised January XX, XXXX.
  	This work was supported by the Grant Agency of the Czech Technical University in Prague SGS16/226/OHK3/3T/13, by the Czech Science Foundation under project No. 15-10280Y, and by the Swedish Foundation for Strategic Research (SSF) under the program Applied Mathematics and the project Complex analysis and convex optimization for EM design.}
\thanks{M.~Capek, V.~Losenicky and L.~Jelinek are with the Department of Electromagnetic Field, Faculty of Electrical Engineering, Czech Technical University in Prague, Technicka 2, 16627, Prague, Czech Republic (e-mail: miloslav.capek@fel.cvut.cz, losenvit@fel.cvut.cz, lukas.jelinek@fel.cvut.cz.).} 
\thanks{M.~Gustafsson is with the Department of Electrical and Information Technology, Lund University, 221~00 Lund, Sweden (e-mail: mats.gustafsson@eit.lth.se).}
}

\markboth{Journal of \LaTeX\ Class Files,~Vol.~XX, No.~X, \today}%
{Capek \MakeLowercase{\textit{et al.}}: Synthetic Benchmark Based on Characteristic Modes}

\maketitle

\begin{abstract}
Characteristic modes of a spherical shell are found analytically as spherical harmonics normalized to radiate unitary power and to fulfill specific boundary conditions. The presented closed-form formulas lead to a proposal of precise synthetic benchmarks which can be utilized to validate the method of moments matrix or performance of characteristic mode decomposition. Dependence on the mesh size, electrical size and other parameters can systematically be studied, including the performance of various mode tracking algorithms. A noticeable advantage is the independence on feeding models. Both theoretical and numerical aspects of characteristic mode decomposition are discussed and illustrated by examples. The performance of state-of-the-art commercial simulators and academic packages having been investigated, we can conclude that all contemporary implementations are capable of identifying the first dominant modes while having severe difficulties with higher-order modes. Surprisingly poor performance of the tracking routines is observed notwithstanding the recent ambitious development.
\end{abstract}


\section{Introduction}
\IEEEPARstart{C}{haracteristic} mode (CM) decomposition~\cite{Garbacz_TCMdissertation, HarringtonMautz_TheoryOfCharacteristicModesForConductingBodies} has become a popular tool for analyzing and designing scatterers and antennas, mainly due to the physical insight gained by modal decomposition without a particular feeding considered~\cite{MartaEva_TheTCMRevisited, VogelEtAl_CManalysis_PuttingPhysicsBackIntoSimulation}. CM decomposition yields a set of real-valued currents which form an orthonormal basis with respect to their radiation patterns and the useful properties of the CMs render this technique appealing for antenna designers~\cite{Martens_2013_MIMO_TCM, YangAdams_SystematicShapeOptimizationOfSymmetricMIMOAntennasUsingCM,EichlerHazdraCapekKorinekHamouz_DesignOfADualBandOrthogonally_AWPL}. Consequently, CM decomposition has been the subject of implementation into commercial tools, such as FEKO~\cite{feko}, \mbox{WIPL-D}~\cite{wipld}, and CST~\cite{cst}, and there also exist a plethora of academic tools, employed primarily for research related to the CM~\cite{atom, CMC, IDA, Makarov_AntennaAndEMModelingWithMatlab}.

The amount of scientific data generated along with the publication activity in the field of CMs is immense. It is therefore surprising that the question of how accurate these results are is scarcely assessed. The rare exception is an early study by Mautz and Harrington, where the results of their FORTRAN implementation~\cite{MauztHarrington_CMprogram} of CM decomposition is compared to the first analytically known eigenvalues of a spherical shell~\cite[Table~II]{HarringtonMautz_ComputationOfCharacteristicModesForConductingBodies}. This lack of detailed numerical benchmarking of the CMs was the main motivation for the developments presented in this paper.

Benchmarking activities, see~\cite{IEEEStd_bench1, IEEEStd_bench2, Warnick2004,WarnickChew_AccuracyOfMoMforScatteringByACylinder} and the references therein for examples, are essential for the validation and quality assessment of methods and tools of computational electromagnetics and are of particular interest for those methods with known numerical issues, as is the case of CM decomposition~\cite{CapekHazdraEichler_AMethodForTheEvaluationOfRadiationQBasedOnModalApproach, SchabEtAl_EigenvalueCrossingAvoidanceInCM, BernabeuValeroVicoKishk_AComparisonBetweenNaturalAndCM, CapekHazdraMasekLosenicky_AnalyticalRepresentationOfCMdec}. As suggested in~\cite{IEEEStd_bench1}, four benchmarks are readily available in computational electromagnetics, namely the comparisons to a closed form solution~\cite{Sihvola_2004, Warnick2004}, to a standard problem~\cite{Helsing_2013, HelsingKarlsson_DeterminationOfNormalizedMagEigenfieldInMwCavities}, to a measurement or to other modeling techniques~\cite{WooWangSchuhSanders_BenchmarkPlateRadarTargets, DeshpandeCockrellBeckNguyen_BenchamarksOfSimpleShapesForLowFreqCEM, Vandenbosch_StateOfTheArtAntennaSWbenchmarking}. The advantages and disadvantages of these possibilities are detailed in~\cite{IEEEStd_bench1} and have led authors to the decision to select a comparison to the analytical model. The drawback of this choice is that analytic solutions are only available for canonical geometries such as ellipsoids and cylinders~\cite{Sihvola_2004, Warnick2004}. The advantage of negligible error levels in the analytic model~\cite{Warnick_NumericalAnalysisForElectromagneticIntegralEquation, JayasekeraCiric_BenchmarkCompFieldLossesForcesForConductingSpheroids}, however, outweighs it.

In this paper we propose four independent benchmarks devoted to various aspects of CM decomposition to validate characteristic eigenvalues, their tracking and conformity between analytically and numerically calculated characteristic currents or characteristic far-field patterns. Moreover, since CMs do not take into account feeding, they can also be used for investigating the accuracy of impedance matrix assemblage which is strongly dependent on discretization~\cite{EichlerHazdraCapek_AspectsOfMeshGenerationTCM}, the selection of basis functions~\cite{PetersonRayMittra_ComputationalMethodsForElectromagnetics}, the quadrature rules used and singularity treatment~\cite{EibertHansen_OnTheCalculationOfPotenticalIntegralsForLinearSourceDistributionsOnTriangularDomains, SieversEibertHansen_TAP2005}. As a testing object we propose a perfectly electrically conducting (PEC) sphere for which the characteristic eigenvalues and characteristic eigencurrents are known analytically~\cite{Garbacz_TCMdissertation}. The symmetry of the spherical shell also introduces eigenspace degeneration~\cite{SchabBernhard_GroupTheoryForCMA} which, together with the null-space of the impedance operator at internal resonances of the shell~\cite{PetersonRayMittra_ComputationalMethodsForElectromagnetics},  introduces serious problems with modal tracking~\cite{CapekHazdraHamouzEichler_AMethodForTrackingCharNumbersAndVectors, LudickJakobusVogel_AtrackingAlgorithmForTheEigenvectorsCalculatedWithCM, MiersLau_WideBandCMtrackingUtilizingFarFieldPatterns, RainesRojas_WidebandCharacteristicModeTracking, SafinManteuffel_AdvancedEigenvalueTrackingofCM}. 

The paper is organized as follows. CM decomposition is briefly recapitulated in Section~\ref{Sec2} and the analytic solution to spherical shell is provided in Section~\ref{Sec3}. The matrix form of CM decomposition is defined in Section~\ref{Sec4}, including a thorough discussion of the numerical issues behind the decomposition. The benchmarks are proposed in Section~\ref{Sec45:benchmarks} and applied on various packages in Section~\ref{Sec5}. The paper is concluded in Section~\ref{Sec7}.

\section{Characteristic Modes Decomposition}
\label{Sec2}

The CMs are introduced \cite{HarringtonMautz_TheoryOfCharacteristicModesForConductingBodies} as solutions to a generalized eigenvalue problem \cite{Wilkinson_AlgebraicEigenvalueProblem} 
\begin{equation}
	\label{eq:CManalyt}
	\mathcal{X}\left(\V{J}_n\right) = \lambda_n\OP{R}\left(\V{J}_n\right)
\end{equation}
where $\mathcal{R}$ and $\mathcal{X}$ represent the real and imaginary parts of the impedance operator \cite{Harrington_FieldComputationByMoM}
\begin{equation}
	\label{eq:Zdef}
	\mathcal{Z}\left(\V{J}_n\right)=\OP{R}\left(\V{J}_n\right)+\J\OP{X}\left(\V{J}_n\right) = \V{\UV{n}} \times \V{\UV{n}} \times \V{E}\left(\V{J}_n\right),
\end{equation}
with $\V{E}$ being the scattered electric field \cite{Harrington_TimeHarmonicElmagField}, $\V{J}_n$ the modal current density, $\lambda_n$ the characteristic eigenvalue and $\V{\UV{n}}$ the unit normal to the PEC surface $\srcRegion$ \cite{Harrington_TimeHarmonicElmagField} which, in this text, coincides with the radial direction. The time-harmonic quantities  under the convention \mbox{$\V{\OP{J}} \left(\V{r},t\right) = \RE\left\{\V{J}\left(\V{r},\omega\right) \mathrm{exp}\left(\J \omega t \right)\right\}$}, with~$\omega$ being the angular frequency, are used throughout the paper. 

The CMs are commonly normalized with respect to unitary radiated power, \ie,
\begin{equation}
\label{eq:CMdef}
\begin{split}
	\frac{1}{2}\int\limits_\srcRegion \V{J}_m^\ast\cdot\mathcal{Z}\left(\V{J}_n\right)\D{S} = &\frac{1}{2}\int\limits_\srcRegion \V{J}_m^\ast\cdot\V{E}\left(\V{J}_n\right)\D{S} \\
    = &\left(1 + \J\lambda_n\right)\delta_{mn} = \kappa_n \delta_{mn},
\end{split}
\end{equation}
where $\delta_{mn}$ is the Kronecker delta \cite{Jeffrey_MathHandbook}, which allows eigenvalues $\lambda_n$ to be expressed as \cite{HarringtonMautz_TheoryOfCharacteristicModesForConductingBodies}
\begin{equation}
	\label{eq:RayleighQuot}
	\lambda_n = \frac{\IM \left\{\kappa_n\right\}}{\RE \left\{\kappa_n\right\}} = \frac{\displaystyle\int\limits_\srcRegion \V{J}_n^\ast\cdot\OP{X}\left(\V{J}_n\right)\D{S}}{\displaystyle\int\limits_\srcRegion \V{J}_n^\ast\cdot\OP{R}\left(\V{J}_n\right)\D{S}},
\end{equation}
in which $\RE\left\{\cdot\right\}$ and $\IM\left\{\cdot\right\}$ denote the real and imaginary parts, respectively. 

Considering only the currents distributed on surfaces, the uniqueness and completeness of CM decomposition is ensured outside internal resonances \cite{ChewSong_GedankenExperimentsToUnderstandInternalResonanceProblem}, \ie{}, when all modal currents $\V{J}_n$ radiate. In the light of \eqref{eq:CManalyt} and \eqref{eq:CMdef}, a sound definition of characteristic modes can thus be stated as follows: \textit{Characteristic modes form a basis of real-valued current densities which diagonalizes the impedance operator $\OP{Z}$ and possess orthonormal radiation patterns.}

\section{Analytical Decomposition}
\label{Sec3}

The analytical solution of CMs decomposition on a spherical shell is presented in this section and the results are to be further used as a reference. We start with a short inspection of systems with potentially known analytical solutions in Section~\ref{Sec3:sep} and the CMs of a spherical shell are presented in Section~\ref{Sec3:sphShell}.

\subsection{Separable Systems}
\label{Sec3:sep}

Orthonormality of far-field radiation patterns and completeness are properties shared between CMs and specific solutions to the vector Helmholtz's equation in separable systems \cite{MorseFeshBach_MethodsOfTheoreticalPhysics}. Particularly, the solutions to the vector Helmholtz's equation \cite{MorseFeshBach_MethodsOfTheoreticalPhysics} in spherical, conical, rectangular and cylindrical (circular-cylindrical, elliptical-cylindrical, paraboloidal-cylindrical) coordinate systems, orthonormalized with respect to the far-field\footnote{Known solutions to the vector Helmholtz's equation in spheroidal coordinates do not guarantee orthonormality \cite{LiKangLeong_SpheroidalWaveFunctionsInEMTheory}.}, can be equated to characteristic modes.

The above-mentioned set of possible candidates is further restricted by a practical requirement on the finite extent of the studied structures so that the model can be discretized without using periodic boundary conditions. Consequently, we are left with two feasible coordinate systems -- spherical and conical. From these two, we have chosen spherical modes othonormal with respect to spherical surfaces \cite{Stratton_ElectromagneticTheory}. Such modes correspond to a surface current density distributed on a spherical shell. In comparison to conical solutions, they are formally simpler and, significantly, exhibit high order degeneracies which complicate modal tracking considerably, see Section~\ref{Sec4}.

\subsection{CM Decomposition of a Spherical Shell}
\label{Sec3:sphShell}

The analytical form of characteristic currents on a spherical shell, see Fig.~\ref{fig1:Sphere}, can be found in the work of Garbacz \cite{Garbacz_TCMdissertation} where it is a result of diagonalization of a scattering matrix, though without any derivation and with the characteristic numbers $\lambda_n$ from \eqref{eq:CManalyt} presented in a slightly different form ($-1/\kappa_n$) which is more favorable for a scattering scenario. Here, instead, we provide a rationale to solve the problem from the perspective of \eqref{eq:CMdef} and then present results which can be used as a standard for numerical tests.

\begin{figure}[t]
\begin{center}
\pgfmathsetmacro{\svec}{0.6}	
\pgfmathsetmacro{\rvec}{.8}
\pgfmathsetmacro{\thetavec}{35}
\pgfmathsetmacro{\phivec}{70}
\def\mFrameRate{10}
\begin{animateinline}[autoplay, loop]{\mFrameRate}
	\multiframe{72}{rt=1+5}{
		\tdplotsetmaincoords{50}{120+\rt}
		\begin{tikzpicture}[scale=4,tdplot_main_coords]
		\shade[tdplot_screen_coords,ball color = white] (0,0) circle (\svec); 
		\shade[tdplot_screen_coords,ball color=blue,opacity=0] (0,0) circle (1.2*\rvec); 
		\coordinate (O) at (0,0,0) node[anchor=east, yshift=5pt]{$\mathbf{0}$};
		\draw[thick,->] (0,0,0) -- (3/4,0,0) node[anchor=north east]{$x$};
		\draw[thick,->] (0,0,0) -- (0,3/4,0) node[anchor=north west]{$y$};
		\draw[thick,->] (0,0,0) -- (0,0,7/8) node[anchor=south]{$z$};
		\tdplotsetcoord{P}{\rvec}{\thetavec}{\phivec}
		\draw[-stealth,color=red,line width=1pt] (O) -- (P) node[above right] {$\boldsymbol{r}$};
		\draw[dashed, color=red] (O) -- (Pxy);
		\draw[dashed, color=red] (P) -- (Pxy);
		\tdplotdrawarc{(O)}{0.2}{0}{\phivec}{anchor=north}{$\varphi$}
		\tdplotsetthetaplanecoords{\phivec}
		\tdplotdrawarc[tdplot_rotated_coords]{(0,0,0)}{0.5}{0}%
		{\thetavec}{anchor=south west}{$\vartheta$}
		\fill[fill=blue,opacity=0.15] (O) -- (0.2,0,0) arc (0:\phivec:0.2);
		\tdplotsetrotatedcoords{-20}{-90}{0}
		\fill[fill=blue,opacity=0.15,tdplot_rotated_coords] (O) -- (0.5,0,0) arc (0:\thetavec:0.5);    
		\end{tikzpicture}
		
	}%
\end{animateinline}
\caption{Sketch of a spherical shell and the used coordinate system.}
\label{fig1:Sphere}
\end{center}
\end{figure}
The orthonormal set of electric fields \cite{Stratton_ElectromagneticTheory}
\begin{equation}
\label{eq:ETE}
\V{E}_{pq}^\mathrm{TE} \left(r \ge a\right) = - \CMC \SPC \V{M}_{pq} \left(\mathrm{h}_p^{(2)}, r,\vartheta,\varphi\right),
\end{equation}
\begin{equation}
\label{eq:ETM}
\V{E}_{pq}^\mathrm{TM} \left(r \ge a\right) = \CMC \SPC \V{N}_{pq} \left(\mathrm{h}_p^{(2)}, r,\vartheta,\varphi\right)
\end{equation}
corresponding to surface current densities
\begin{equation}
\label{eq:JTE}
\V{J}_{pq}^\mathrm{TE} = \CMC \UV{\V{r}}\times \V{N}_{pq} \left(\mathrm{j}_p,a,\vartheta,\varphi\right),
\end{equation}
\begin{equation}
\label{eq:JTM}
\V{J}_{pq}^\mathrm{TM} = \CMC \UV{\V{r}}\times \V{M}_{pq} \left(\mathrm{j}_p,a,\vartheta,\varphi\right),
\end{equation}
distributed on a spherical shell of radius $r=a$ are the desired solutions to the vector Helmohltz's equation for $\V{r}$ denoting radial direction,
\begin{equation}
\label{eq:gamma}
\SPC = \ZVAC ka \, \JBSL{p}{ka} \frac{\partial \left( ka \, \JBSL{p}{ka} \right)}{\partial ka},
\end{equation}
functions \mbox{$\V{M}_{pq}$} and \mbox{$\V{N}_{pq}$} being defined in \cite{Stratton_ElectromagneticTheory}, \mbox{$\ZVAC = \sqrt{\MUE/\EPS}$} being the free-space impedance and $\mathrm{j}_p$ ($\mathrm{h}_p^{(2)}$) being the spherical Bessel (Hankel) function of $p$-th order \cite{Jeffrey_MathHandbook}. Setting then 
\begin{equation}
\label{eq:C}
\CMC = \frac{k}{\SPC} \displaystyle\sqrt{\frac{\ZVAC\left(2p+1\right)\left(p-q\right)!}{\pi \left( 1 + \delta_{q0} \right) p \left(p+1\right)\left(p+q\right)!}},
\end{equation}
such solutions also satisfy~\eqref{eq:CMdef} and can be identified with the characteristic modes of a spherical shell, see Fig.~\ref{fig1:Sphere}. Substituting \eqref{eq:ETE}, \eqref{eq:ETM}, \eqref{eq:JTE} and \eqref{eq:JTM} into \eqref{eq:CMdef}, the characteristic numbers are found in analytic form as
\begin{equation}
\label{eq:lambdaTE}
\lambda_p^\mathrm{TE} =-\frac{\YBSL{p}{ka}}{\JBSL{p}{ka}} 
\end{equation}
and
\begin{equation}
\label{eq:lambdaTM}
\lambda_p^\mathrm{TM} = -\frac{\left(p+1\right) \YBSL{p}{ka} - ka \, \YBSL{p+1}{ka}}{\left(p+1\right) \JBSL{p}{ka} - ka \, \JBSL{p+1}{ka}},
\end{equation}
where $\mathrm{y}_p$ is the spherical Bessel function of the second kind and $p$-th order.

\begin{figure}[t]
	\begin{center}
		\includegraphics[width=\figwidth cm]{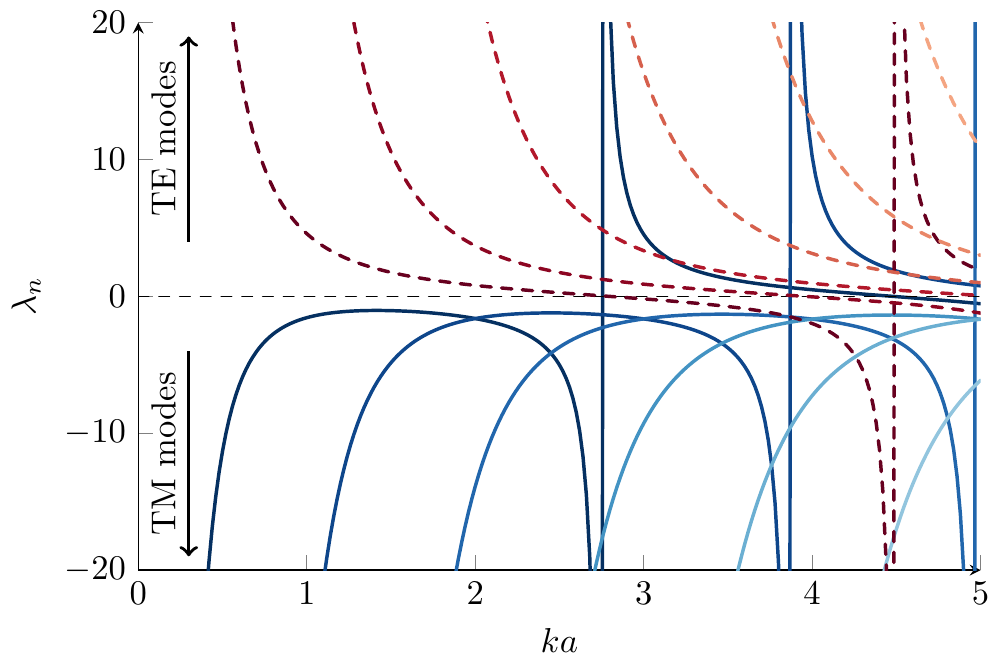}
		\caption{Characteristic eigenvalues $\lambda_n$ of a spherical shell of radius $a$. The first six TE and TM modes are depicted. The vertical lines (where the eigenvalues are not well-defined functions) correspond to the internal resonances and they are kept in the figure to simplify the tracking of different modes. Modes with \mbox{$\lambda_n > 0$} are predominantly inductive while \mbox{$\lambda_n < 0$} are predominantly capacitive.}
		\label{fig3:SphereAnalyticNumbers}
	\end{center}
\end{figure}

\begin{figure}[t]
	\begin{center}
		\includegraphics[width=\figwidth cm]{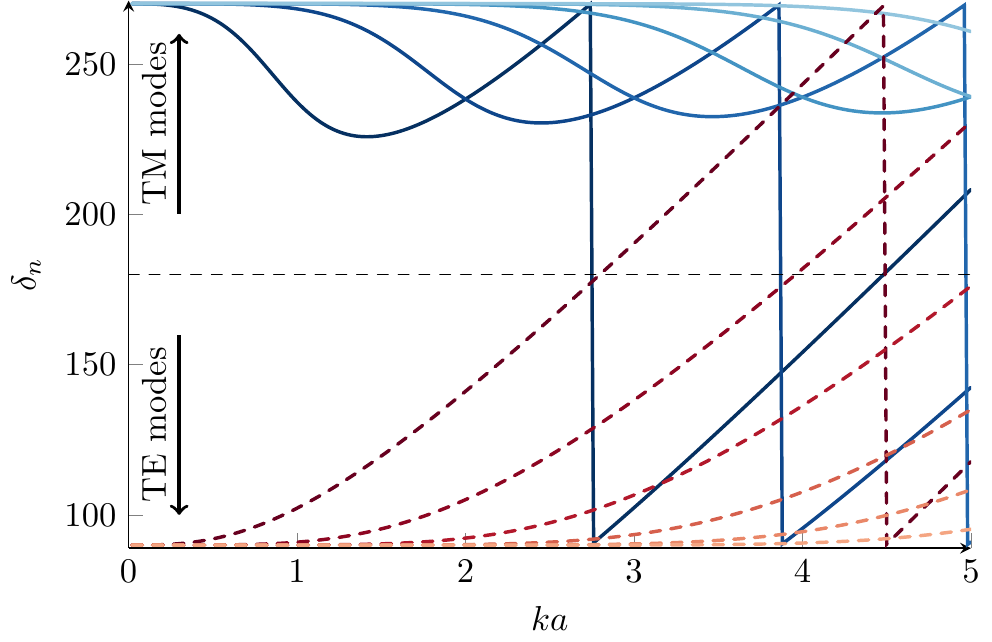}
		\caption{Characteristic eigenangles $\delta_n$ of a spherical shell. The first six TM and TE modes are depicted, \cf{} Fig.~\ref{fig3:SphereAnalyticNumbers}.}
		\label{fig4:SphereAnalyticAngles}
	\end{center}
\end{figure}

To simplify the notation, an aggregated index $n$ is adopted from \cite{Hansen_SphericalNearFieldAntennaMeasurements} as
\begin{equation}
\label{eq:aggIndex}
\V{J}_n = \left\{ \V{J}_{pq}^\mathrm{TE}, \V{J}_{pq}^\mathrm{TM} \right\},
\end{equation}
and the characteristic numbers, \mbox{$\lambda_n \in \left(-\infty,\infty\right)$}, are rescaled in terms of so-called characteristic angles \mbox{$\delta_n \in \left[90\degree, 270\degree\right]$} as \cite{Newman_SmallAntennaLocationSynthesisUsingCharacteristicModes}
\begin{equation}
\label{eq:charAngles}
\delta_n = 180 \left( 1 - \frac{1}{\pi} \arctan\left(\lambda_n\right) \right).
\end{equation}
The characteristic numbers $\lambda_n$ belonging to the first six TE and TM modes (not counting degenerations) are depicted in Fig.~\ref{fig3:SphereAnalyticNumbers} and the corresponding characteristic angles $\delta_n$ are depicted in Fig.~\ref{fig4:SphereAnalyticAngles}.

\section{Numerical Evaluation}
\label{Sec4}

A numerical solution to the characteristic modes of a spherical shell is found in this section, including a discussion of related numerical issues.

In the common treatment \cite{PetersonRayMittra_ComputationalMethodsForElectromagnetics}, the operator \eqref{eq:Zdef} is represented in a basis of piecewise functions $\left\{\basfcn_u\right\}$ \cite{PetersonRayMittra_ComputationalMethodsForElectromagnetics} in the form of the impedance matrix
\begin{equation}
\label{eq:discret1}
\M{Z} = \M{R} + \J\M{X} = \left[Z_{uv}\right] = \left[\,\int\limits_\srcRegion \basfcn_u^\ast \OP{Z} \left(\basfcn_v\right) \D{S} \right]
\end{equation}
and the relation \eqref{eq:CManalyt} is transformed into the matrix equation
\begin{equation}
\label{eq:CMnum}
\M{X} \Ivec_n = \lambda_n \M{R} \Ivec_n
\end{equation}
with modal current density from \eqref{eq:CManalyt} calculated as
\begin{equation}
\label{eq:discret2}
\M{J}_n = \sum\limits_u I_{nu} \basfcn_u.
\end{equation}
Formula \eqref{eq:CMnum} is of general validity, and, therefore, characteristic modes of arbitrarily shaped bodies can be found in this way at the expense, however, of the occurrence of various numerical issues and artifacts which are discussed below.

\subsection{Known Numerical Issues}
\label{Sec4:issues}

\begin{figure}[t]
	\begin{center}
		\includegraphics[width=9cm]{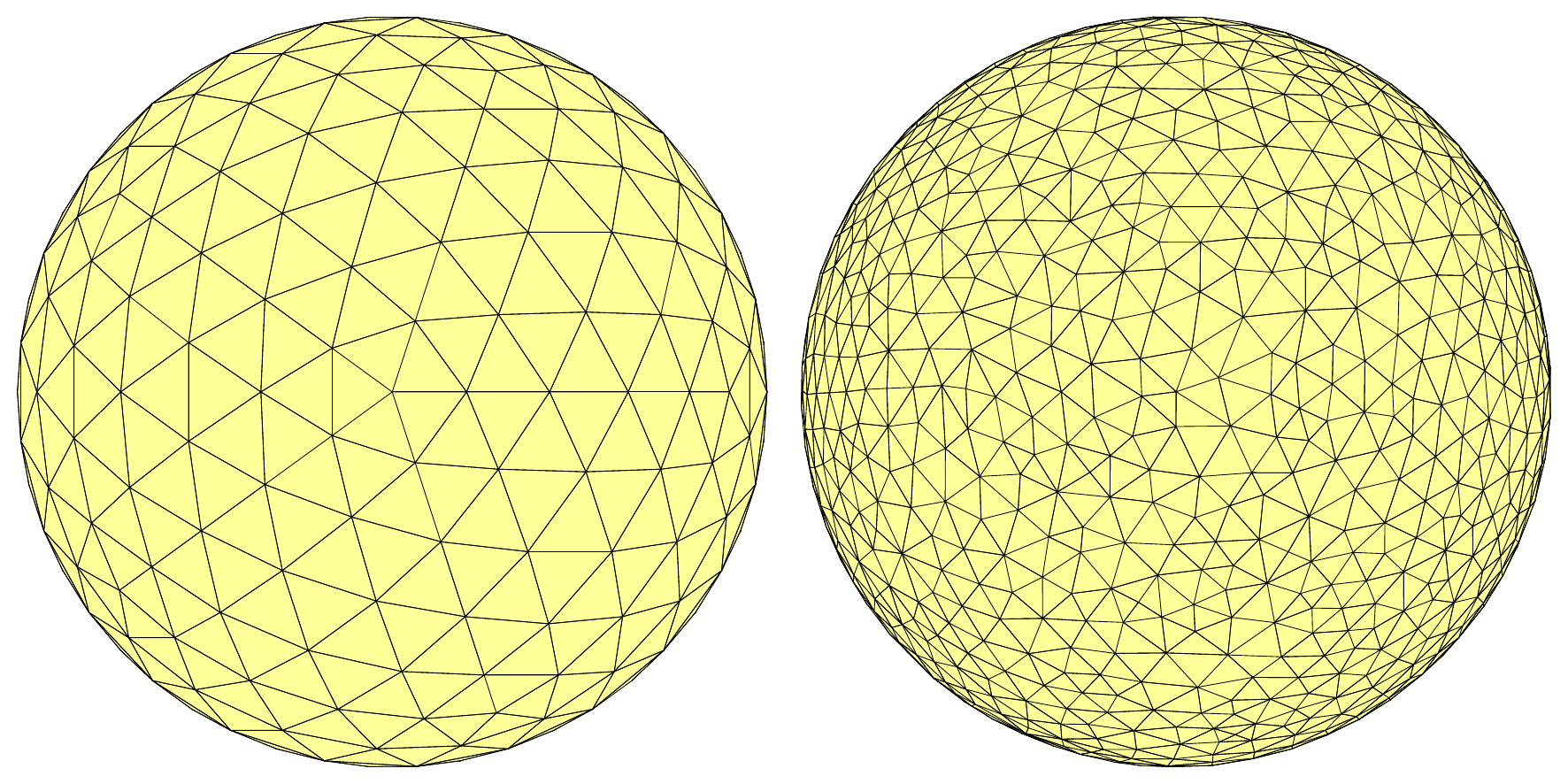}
		\caption{Spherical shell from Fig.~\ref{fig1:Sphere} triangularized into $500$~triangles (left) and $2200$~triangles (right) with $750$~(left) and $3300$~(right) RWG basis functions, respectively. The coarser discretization is used throughout the paper and, wherever required, the results are compared with those for the denser grid.}
		\label{fig5:SphereNumericalMesh}
	\end{center}
\end{figure}

The numerical issues connected to \eqref{eq:CMnum} will be demonstrated on a spherical shell discretized into a triangular mesh grid, see Fig.~\ref{fig5:SphereNumericalMesh}, with RWG basis functions \cite{RaoWiltonGlisson_ElectromagneticScatteringBySurfacesOfArbitraryShape} applied. The mesh grid was exported from FEKO \cite{feko} as a NASTRAN file \cite{nastran} and is freely available \cite{CMASphericalShell_benchWeb}. Spherical geometry allows a near-perfect triangular mesh consisting of equiangular triangles to be generated, see Fig.~\ref{fig5:SphereNumericalMesh}.

When dealing with the CMs, the following issues arise:

\subsubsection{Indefiniteness of a Real Part of the Impedance Matrix}

The importance of algebraic properties of matrices~$\M{R}$ and~$\M{X}$ can readily be seen from~\eqref{eq:CMdef} and \eqref{eq:CMnum}. To obtain real characteristic numbers and vectors, these matrices have to be symmetric, \ie{}, Galerkin testing procedure should properly be applied~\cite{Harrington_FieldComputationByMoM}. If matrices are slightly non-symmetric, they can be symmetrized \textit{ex post} as
\begin{equation}
\label{eq:Rsymmetr}
\M{Z}_\mathrm{sym} = \frac{1}{2} \left( \M{Z} + \M{Z}^\trans\right),
\end{equation}
in which superscript $\trans$ denotes matrix transposition. Since matrix $\M{R}$ serves in \eqref{eq:CMnum} as a weighing operator \cite{Wilkinson_AlgebraicEigenvalueProblem} and represents radiated power, its positive definiteness, \mbox{$\M{R} \succ 0$}, is crucial. In fact, a potential violation of this condition is one of the biggest known issue related to the CMs \cite{HarringtonMautz_ComputationOfCharacteristicModesForConductingBodies}. On the contrary, matrix~$\M{X}$ is usually well-posed.

\begin{figure}[t]
	\begin{center}
		\includegraphics[width=\figwidth cm]{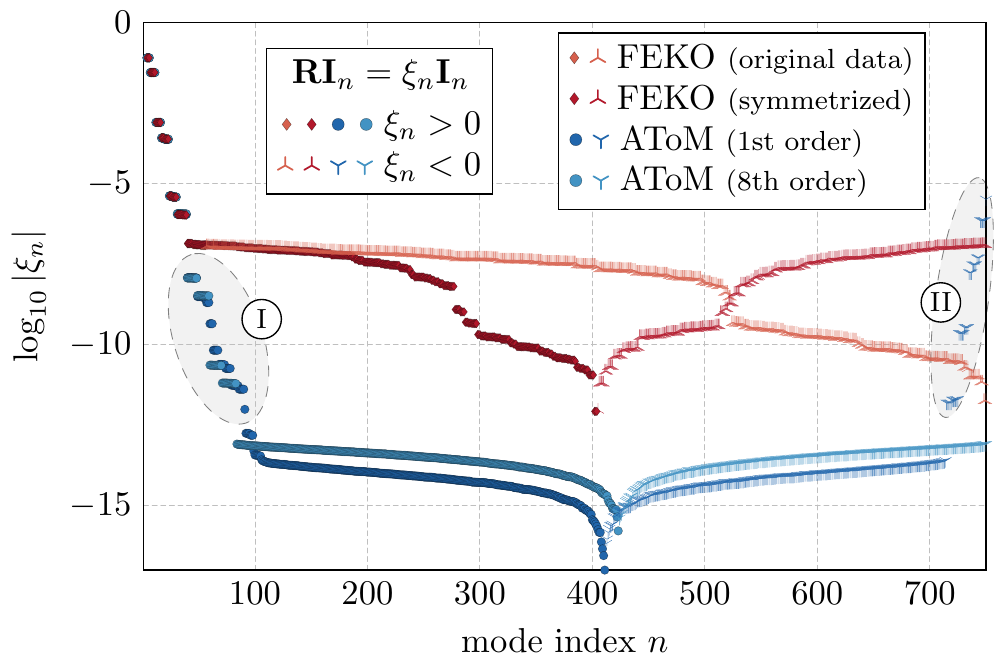}
		\caption{Eigenvalues of a real part of the impedance matrix of a spherical shell discretized into~$500$ triangles.}
		\label{fig:SphereReigNumbers}
	\end{center}
\end{figure}

In order to present this last problem numerically, matrix $\M{R}$ of a spherical shell discretized into 500~triangles (left panel in Fig.~\ref{fig5:SphereNumericalMesh}) is decomposed as
\begin{equation}
\label{eq:Rdecomp}
\M{R} \hat{\Ivec}_n = \xi_n \hat{\Ivec}_n
\end{equation}
and the eigenvalues~$\xi_n$ are depicted in Fig.~\ref{fig:SphereReigNumbers}. The results correspond to matrix~$\M{R}$ obtained by two different packages: the commercial package FEKO~\cite{feko} and the academic tool AToM~\cite{atom}. For the sake of clarity, the absolute values of the eigenvalues are shown in logarithmic scale and the negativity of the eigenvalues is indicated by the marker (star) used. Note that eigenvalues~$\xi_n$ are proportional to the radiated power \cite{Harrington_FieldComputationByMoM}.  

Generally, we see that only a few modes radiate well (top left part of Fig.~\ref{fig:SphereReigNumbers}) and that at least one half of all eigenvalues are negative (right half of Fig.~\ref{fig:SphereReigNumbers}). These negative eigenvalues are related to the presence of ubiquitous numerical noise and they should be removed from matrix~$\M{R}$ using \eqref{eq:Rdecomp} and substituting \mbox{$\xi_n \equiv 0$} \cite{GustafssonTayliEhrenborgEtAl_AntennaCurrentOptimizationUsingMatlabAndCVX} in the consecutive back-composition
\begin{equation}
\label{eq:Zcomp}
\M{R}_\mathrm{pos} = \hat{\Ivec} \V{\xi} \hat{\Ivec}^\trans,
\end{equation}
where $\hat{\Ivec}$ is a matrix containing column vectors $\hat{\Ivec}_n$ and $ \V{\xi}$ is a diagonal matrix containing eigenvalues $ \xi_n$. Unfortunately, formula \eqref{eq:Zcomp} cannot cure the imminent fact that matrix~$\M{R}$ is ill-conditioned.

Matrix $\M{R}$ originating from FEKO is non-symmetric, which causes its different appearance as compared to other curves. For the second set of eigenvalues, the matrix has been symmetrized according to \eqref{eq:Rsymmetr} prior to decomposition \eqref{eq:Rdecomp}. Two different orders of Gaussian quadrature rule \cite{Dunavant_HighDegreeEfficientGQR} have been used in AToM \cite{Harrington_FieldComputationByMoM}. In comparison to FEKO, AToM has been able to find more modes, namely those highlighted in the shaded oval denoted by number~I. This fact will play an important role at a further point and is most probably caused by single precision arithmetic used in FEKO, see Appendix~\ref{appFEKO} for the simulation setup. It is also demonstrated by high negative eigenvalues highlighted by oval~II in Fig.~\ref{fig:SphereReigNumbers} that centroid approximation (only one quadrature point in each triangle) does not lead to a well behaved matrix $\M{R}$. It is obvious that the number of properly found modes directly reflect the quality of numerics used when constructing the impedance matrix.

\subsubsection{Mode Degeneracy}

Degeneracy can be traced back to the geometrical symmetries of the $\srcRegion$ region \cite{LandauLifshitz_QuantumMechanicsVol3, Knorr_1973_TCM_symmetry, SchabBernhard_GroupTheoryForCMA} and it poses complications with mode tracking \cite{CapekHazdraHamouzEichler_AMethodForTrackingCharNumbersAndVectors, SafinManteuffel_AdvancedEigenvalueTrackingofCM, SchabEtAl_EigenvalueCrossingAvoidanceInCM}. A spherical shell has a degeneracy \cite{Stratton_ElectromagneticTheory} of
\begin{equation}
\label{eq:degener}
N \left( p \right) = 2 p + 1,
\end{equation}
which means that for $p$-th order there are $N_p$ degenerated modes. Therefore, each solid and dashed line in Fig.~\ref{fig3:SphereAnalyticNumbers} and Fig.~\ref{fig4:SphereAnalyticAngles} has $2 p + 1$ multiplicity. The effect of degeneracy is also seen in Fig.~\ref{fig:SphereReigNumbers} as it appears as stairs spreading their length from left to right. Except for geometrical degeneracies~\eqref{eq:degener}, the numerical tracking procedure is yet more complicated for the occasional degeneracies occurring at frequencies where TM and TE modes intersect.

The possibility of reducing the number of geometrically degenerated modes lies in the utilization of the procedure from \cite{Knorr_1973_TCM_symmetry} which, however, relies on a particular choice of basis functions and it is therefore problem-dependent. One useful work-around is to make the mesh grid markedly non-symmetrical which helps to remove the degeneracies.

\subsubsection{Internal Resonances}

Internal resonances are inherent to all closed PEC surfaces \cite{ChewSong_GedankenExperimentsToUnderstandInternalResonanceProblem, SarkarMokoleSalazarPalma_AnExposeOnInternalResonancesCM, PetersonRayMittra_ComputationalMethodsForElectromagnetics} and make the matrix $\M{Z}$ ill-conditioned at resonance frequencies. They occur at those frequencies at which non-radiating current sources may exist, \ie{}, where denominators of \eqref{eq:lambdaTE} and \eqref{eq:lambdaTM} vanish. The consequences of solving \eqref{eq:CMnum} at these frequencies are explained in \cite{SarkarMokoleSalazarPalma_AnExposeOnInternalResonancesCM,Jimenez_2013_TCM_sphere, BernabeuValeroVicoKishk_AComparisonBetweenNaturalAndCM}, where it is also important to notice the comparison between resonances of the CMs and natural modes. At the internal resonance, the characteristic numbers abruptly change value between plus and minus infinity, see Fig.~\ref{fig3:SphereAnalyticNumbers}. The majority of tested packages suffer from this issue, although they can effectively be solved using Combined Field Integral Equation (CFIE) \cite{DaiLiuGanChew_CFIEforCMs}.

\subsubsection{Dependence on Conformity and Mesh Density}

Equation \eqref{eq:CMnum} is solved for approximate models like those depicted in Fig.~\ref{fig5:SphereNumericalMesh} and the question arises of which mesh scheme describes the original smooth object best \cite{PetersonRayMittra_ComputationalMethodsForElectromagnetics, Warnick_NumericalAnalysisForElectromagneticIntegralEquation}. Quality of mesh grid, mesh density with respect to conformity, rounding errors, and computational requirements should all be taken into account when dealing with this question. It is also important to note that, although a finer mesh better describes the original object, the increased number of potential modes is redeemed by a higher level of numerical noise resulting in more invalid modes, \cf{} Fig.~\ref{fig:SphereReigNumbers}, longer computational time and higher memory consumption.

\subsubsection{Dependence on Integration and Singularity Treatment}

\begin{figure}[t]
	\begin{center}
		\includegraphics[width=\figwidth cm]{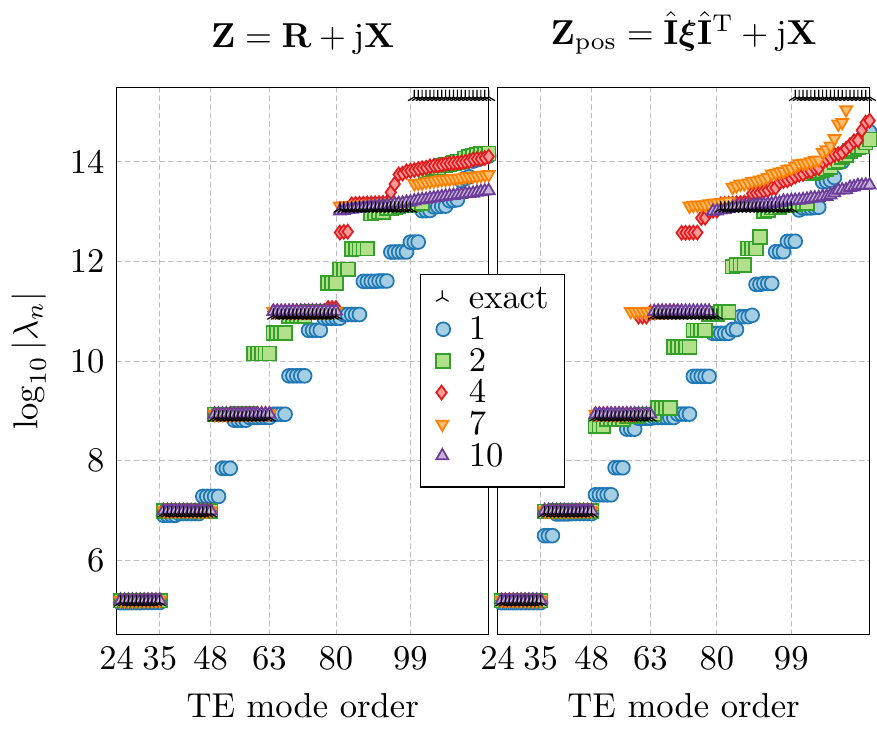}
		\caption{Comparison of characteristic eigenvalues of the impedance matrix found by the AToM package with analytically known results. The results correspond to a spherical shell discretized into 500~triangles with electric size \mbox{$ka = 1.5$}. Various orders of quadrature rule \cite{Dunavant_HighDegreeEfficientGQR} used in the evaluation of the impedance matrix were used and the graph is zoomed in on the area in which the data differ the most. The left panel shows CM decomposition \eqref{eq:CMnum} of the original impedance matrix $\M{Z}$ \eqref{eq:discret1}, while the right panel shows CM decomposition of the modified impedance matrix $\M{Z}_\mathrm{pos}$ with a positively definite real part \eqref{eq:Zcomp}.}
		\label{fig15:SphereAnalyticNumbersAbsLog}
	\end{center}
\end{figure}

As with the previous issue, this point is merely a technicality, yet it strongly influences the final results. The higher order quadrature rules~\cite{PetersonRayMittra_ComputationalMethodsForElectromagnetics} have a great impact on the quality of results, see Fig.~\ref{fig15:SphereAnalyticNumbersAbsLog}. Higher-order basis functions can be advantageously applied as well \cite{PetersonRayMittra_ComputationalMethodsForElectromagnetics}. Special care should also be taken with singularities \cite{EibertHansen_OnTheCalculationOfPotenticalIntegralsForLinearSourceDistributionsOnTriangularDomains, SieversEibertHansen_TAP2005} occurring during the evaluation of \eqref{eq:discret1}.

\section{Benchmarks}
\label{Sec45:benchmarks}

Knowing the analytic results and common issues behind characteristic mode decomposition, the following benchmarks are proposed:
\begin{description}[leftmargin=!,labelwidth=\widthof{\bfseries Test~\#1}]
	\item[Test~\#1] Characteristic eigenvalues of impedance matrix~$\M{Z}$ for a given electrical size~$ka$, see Section~\ref{Bench1}.
	\item[Test~\#2] Modal tracking in a given range of~$ka$, see Section~\ref{Bench2}.
	\item[Test~\#3] Conformity between analytically and numerically calculated characteristic currents~$\V{J}_n$, see Section~\ref{Bench3}.
  	\item[Test~\#4] Correspondence of analytically and numerically calculated characteristic far-fields, see Section~\ref{Bench4}.
\end{description}

All the aforementioned tests are extremely simple to implement and add minimal demands on post-processing. The key features to be investigated within all the tests are
\begin{itemize}
	\item choice of the basis and testing functions,    
	\item precision of the integration scheme used,
	\item precision and robustness of the singularity treatment used,
	\item quality of the tracking algorithm.
\end{itemize}
Notice that a fixed mesh is used and that the errors due to meshing are not considered.

\subsection{Test~\#1}
\label{Bench1}

\begin{figure}[t]
	\begin{center}
		\includegraphics[width=\figwidth cm]{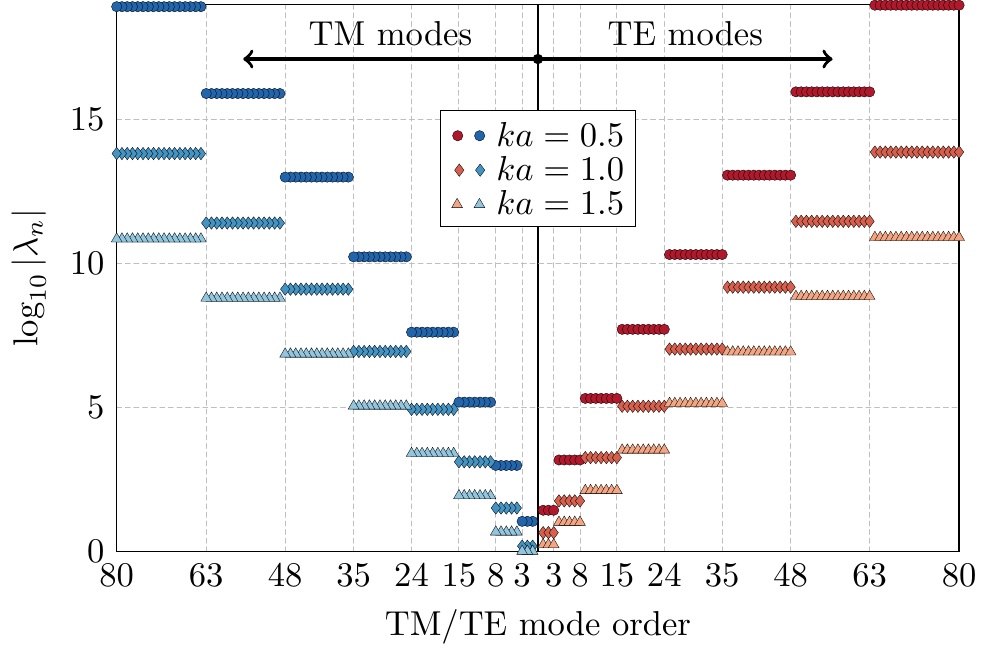}
		\caption{Analytically evaluated characteristic numbers of TM and TE modes of a spherical shell for several electrical sizes $ka$.}
		\label{fig2:SphereAnalyticNumbersAbsLog}
	\end{center}
\end{figure}

The first test focuses on the quality of an impedance matrix, thus making it an efficient benchmark of the method of moments codes. The analytically known eigenvalues $\lambda_n$ from \eqref{eq:lambdaTE} and \eqref{eq:lambdaTM} are compared with eigenvalues obtained by CM decomposition~\eqref{eq:CMnum} of impedance matrix~$\M{Z}$ from~\eqref{eq:discret1}, its symmetrized form~$\M{Z}_\mathrm{sym}$ from \eqref{eq:Rsymmetr}, or of an impedance matrix with a positively definite real part \eqref{eq:Zcomp}. The test is proposed for electrical sizes~\mbox{$ka=0.5$} and~\mbox{$ka=1.5$}, so that both electrically small and reasonably large objects are tested. The spectrum of eigenvalues is calculated by the generalized Schur decomposition \cite{Wilkinson_AlgebraicEigenvalueProblem} (\texttt{eig} with QZ algorithm in Matlab) and all eigenvalues which are infinite, complex-valued or correspond to mode with \mbox{$\Ivec_n^\herm \M{R} \Ivec_n \leq 0$} are removed. The analytical results are depicted in Fig.~\ref{fig2:SphereAnalyticNumbersAbsLog}. The maximum number of modes for three different numerical precisions are summarized in Table~\ref{tab:NumPrecision}.

\begin{table}
	\begin{center}
		\begin{tabular}{ccccc}
			\toprule
			\multirow{3}*{$ka$} &  & Single p. & Double p. & Quadruple p. \\ \cmidrule{2-5}
			& bits~$b$ & $23$ & $52$ & $112$ \\ \cmidrule{2-5} 
			& $\log_{10}\left(2^b\right)$ & $\approx 6.924$ & $\approx 15.65$ & $\approx 33.72$ \\ [2pt]\toprule
			\multirow{2}*{$0.5$} & $p_\mathrm{max}^\mathrm{TE/TM}$ & $4/4$ & $7/7$ & $12/12$ \\ \cmidrule{2-5} 
			& \mbox{$N \left(p_\mathrm{max}^\mathrm{TE/TM}\right)$} & $24/24$ & $63/63$ & $168/168$ \\ \toprule
 			\multirow{2}*{$1.5$} & $p_\mathrm{max}^\mathrm{TE/TM}$ & $6/6$ & $10/10$ & $17/17$ \\ \cmidrule{2-5} 
			& \mbox{$N \left(p_\mathrm{max}^\mathrm{TE/TM}\right)$} & $48/48$ & $120/120$ & $323/323$ \\ \bottomrule
		\end{tabular}
	\end{center}
	\caption{Number of theoretically achievable TE and TM modes dependent on used numerical precision in Matlab and electrical size~$ka$. The number of TE and TM modes $N$ was calculated according to \eqref{eq:degener} with~$p_\mathrm{max}^\mathrm{TE}$ and~$p_\mathrm{max}^\mathrm{TM}$ substituted.}
\label{tab:NumPrecision}    
\end{table}

\subsection{Test~\#2}
\label{Bench2}

The second test investigates the frequency behavior of characteristic eigenvalues. The setup is as follows: 100~modes should be calculated at 226~equidistantly spaced frequency samples between~\mbox{$ka = 0.5$} and~\mbox{$ka = 5$}, and mode tracking should be provided. The number of modes is selected from Fig.~\ref{fig3:SphereAnalyticNumbers}, considering mode degeneracy \eqref{eq:degener} so that, theoretically, all required data to track the first six TE and TM modes are provided (\mbox{$2 \sum_1^6 \left( 2p+1 \right) = 96$}). The frequency span is chosen to cover a couple of internal resonances. The number of frequency samples is chosen as a trade-off between sufficient~$ka$ sampling \mbox{($\Delta ka = 0.02$}) and a computationally feasible solution. The reference solution is depicted in Fig.~\ref{fig3:SphereAnalyticNumbers} and Fig.~\ref{fig4:SphereAnalyticAngles}.

\subsection{Test~\#3}
\label{Bench3}

The conformity of numerically calculated characteristic currents~$\M{J}_n$ with the analytical results \eqref{eq:JTE} and \eqref{eq:JTM} is studied with the  third test using
\begin{equation}
\label{eq:JJtestN}
\chi_n = \displaystyle \max_p \sqrt{\sum\limits_{q=-p}^{p} \ABS{ \int\limits_\srcRegion \M{\hat{J}}_n \left(\M{r}\right) \cdot \M{\hat{J}}_{pq}^\mathrm{TM/TE} \left(\M{r}\right) \D{S}}^2},
\end{equation}
and the current \mbox{$\M{\hat{J}} \left(\M{r}\right)$} is normalized according to
\begin{equation}
\label{eq:JJnorm}
\M{\hat{J}} = \displaystyle\frac{\M{J}}{\displaystyle\sqrt{\int\limits_\srcRegion \M{J} \left(\M{r}\right)\cdot \M{J} \left(\M{r}\right) \D{S}}}.
\end{equation}
Ideally, the coefficient~$\chi_n$ should be equal to unity for all~$n$.

\subsection{Test~\#4}
\label{Bench4}

Far-fields \mbox{$\V{F}_{pq}^\mathrm{TE/TM} \left(\vartheta,\varphi\right)$} of analytical characteristic modes of a spherical shell can be deduced from \eqref{eq:ETE}, \eqref{eq:ETM} and, following the definition of characteristic modes from Section~\ref{Sec2}, they should form an orthogonal and complete set. Since characteristic radiation patterns are commonly used in practice~\cite{MiersLau_WideBandCMtrackingUtilizingFarFieldPatterns, SafinManteuffel_ReconstructionofCMonAntenna}, a meaningful test is to compare numerically evaluated characteristic far-fields~$\V{F}_n$ with analytical ones via 
\begin{equation}
\label{eq:test3eq6}
\zeta_n = \displaystyle \frac{\max\limits_p \displaystyle \sum\limits_{q=-p}^{p} \left| P_{pq,n}^\mathrm{TE/TM} \right|^2}{\displaystyle \sum\limits_{p,q} \sum\limits_\mathrm{TE/TM}\left| P_{pq,n}^\mathrm{TE/TM} \right|^2}
\end{equation}
where \begin{equation}
\label{eq:test3eq4}
P_{pq,n}^\mathrm{TE/TM} = \displaystyle \frac{1}{2 \ZVAC} \int\limits_{4\pi} \left( \V{F}_{pq}^\mathrm{TE/TM} \right)^\ast \cdot \V{F}_n \sin \vartheta \D{\vartheta} \D{\varphi}.
 \end{equation}
The metric $\zeta_n$ should ideally be equal to unity for all~$n$.

As for the previous test, metric~$\zeta_n$ should mostly judge the quality of numerically evaluated current patterns which are directly reflected in the corresponding far-fields. Simultaneously, it tests the fundamental property of CMs, far-field orthogonality.

\section{Results}
\label{Sec5}

This section presents the results of four tests from the last section performed on the 	numerical packages implementing method of moment solution to field integral equations, namely on:
\begin{itemize}
	\item FEKO \cite{feko},
	\item CST-MWS \cite{cst},
	\item WIPL-D \cite{wipld},
	\item CEM One \cite{CemOne}
	\item AToM \cite{atom},
	\item Makarov \cite{Makarov_AntennaAndEMModelingWithMatlab}.
\end{itemize}
The most important settings of the solvers used are specified in Appendices~\ref{appFEKO}--\ref{appAToM}.

\subsection{Test~\#1}
\label{Results1}

\begin{figure}[t]
	\begin{center}
		\includegraphics[width=\figwidth cm]{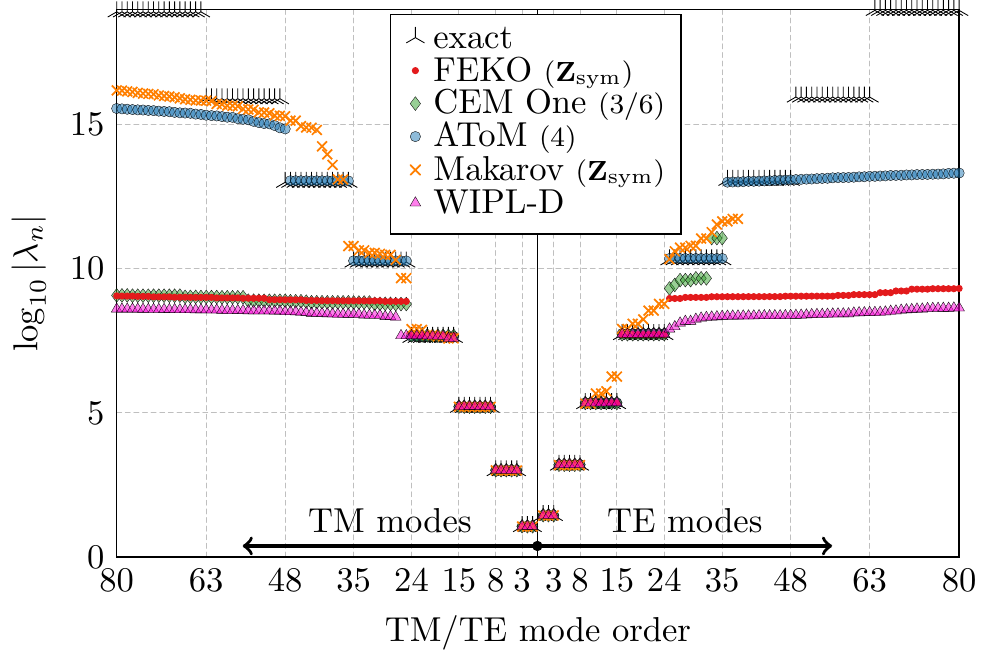}
		\caption{Characteristic numbers for a spherical shell of electrical size \mbox{$ka = 0.5$} discretized into $500$~triangles. Selected commercial and in-house tools are compared. The small number in brackets means the order of quadrature rule. Matrices which have been manually symmetrized are explicitly mentioned.}
		\label{fig6:numSFLam05}
	\end{center}
\end{figure}

\begin{figure}[t]
	\begin{center}
		\includegraphics[width=\figwidth cm]{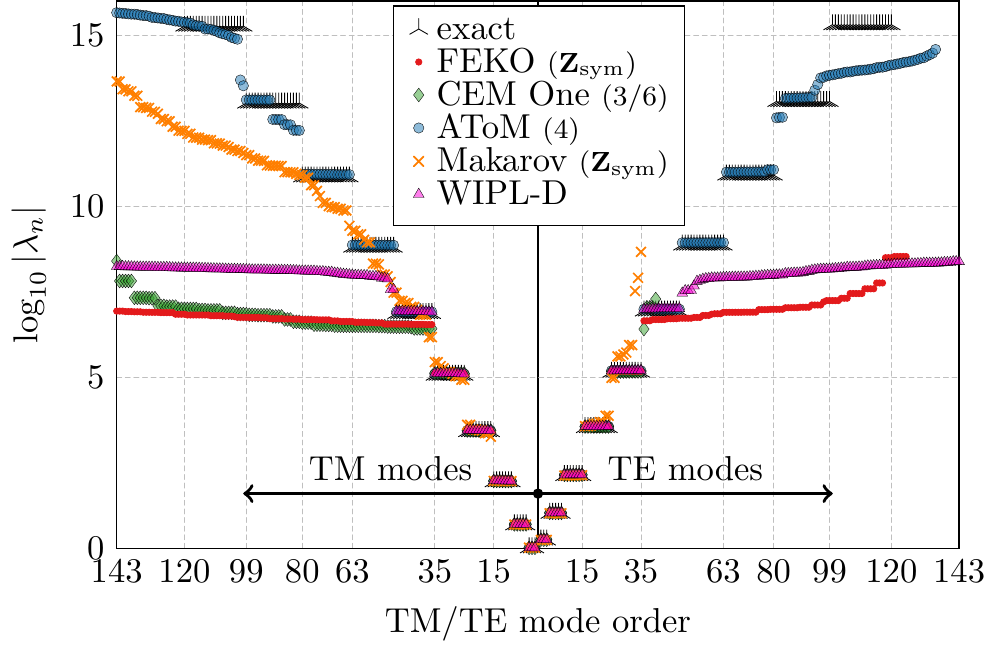}
		\caption{Characteristic numbers for a spherical shell of electrical size \mbox{$ka = 1.5$} discretized into $500$~triangles. The meaning of all abbreviations is the same as in Fig.~\ref{fig6:numSFLam05}.}
		\label{fig7:numSFLam05}
	\end{center}
\end{figure}

The results of the first test are depicted in Fig.~\ref{fig6:numSFLam05} for electric size \mbox{$ka=0.5$} and in Fig.~\ref{fig7:numSFLam05} for electric size \mbox{$ka=1.5$}, respectively. In both cases, the spherical shell was discretized into $500$~discretization elements and, unless otherwise stated, the triangular mesh grid from Fig.~\ref{fig5:SphereNumericalMesh} was used. The impedance matrices of FEKO and Makarov's code are originally slightly non-symmetric, therefore \eqref{eq:Rsymmetr} was applied before the analysis. The order of Gaussian quadrature in CEM One and AToM can be controlled by the user which is why the selected integration scheme is explicitly mentioned in the parentheses. Data from CST-MWS were not analyzed\footnote{The impedance matrix is not accessible and it is thus not guaranteed that the same algorithm for eigenvalue decomposition is used for all tested cases.}.

The accuracy of the academic package AToM is limited by the double numerical precision used. Therefore, the six and eight lowest order non-degenerated TE and TM modes are correctly found in Fig.~\ref{fig6:numSFLam05} and Fig.~\ref{fig7:numSFLam05}, respectively. The commercial packages are able to represent the first four or five modes, most likely for the use of single precision. Notice that, theoretically, $14$ or $20$ non-degenerated TM and TE modes could be retrieved, see Table~\ref{tab:NumPrecision}. In the light of this, the number of correctly found modes is relatively low. The lack of numerical precision can be partially compensated using VPA (Variable Precision Arithmetic) in Matlab or the Multiprecision Computing Toolbox~\cite{advanpix}. However, these improvements have the potential to add only a couple of additional modes and always at the cost of extreme computational time.

The presented results suggest that the slight non-symmetry of an impedance matrix is not a major issue. However, care should be taken with the numerical precision used and the application of the high-order quadrature rule in conjunction with the precise implementation of singular terms.

\subsection{Test~\#2}
\label{Results2}

This test focuses on the mode tracking algorithm, therefore, only packages with this utility implemented could be analyzed. In all cases, the tracking analysis was performed in the frequency range from~\mbox{$ka=0.5$} to \mbox{$ka=5$}, and then, to produce lucid graphical outputs, only the problematic region of~\mbox{$ka\in\left[2.5,5\right]$} was depicted.

\begin{figure}[t]
	\begin{center}
		\includegraphics[width=\figwidth cm]{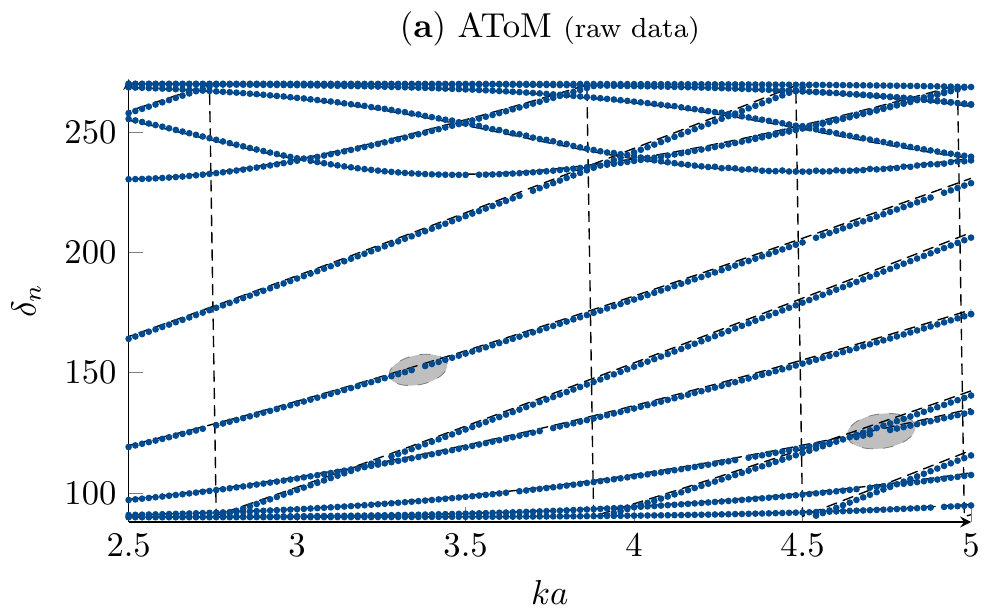}
		
		\vspace{0.3cm}		
		\includegraphics[width=\figwidth cm]{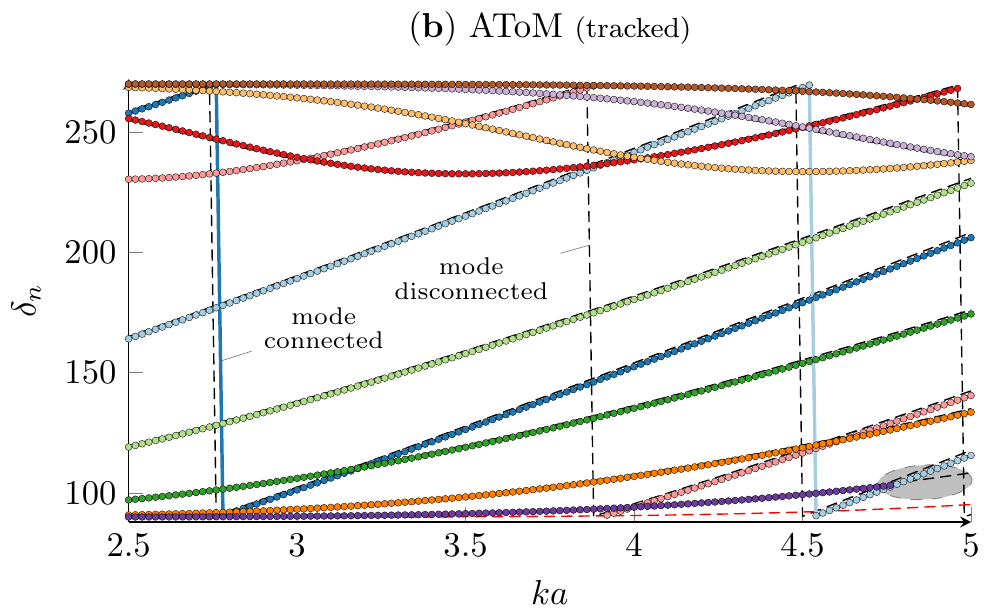}				
		\caption{Test of tracking procedure provided by the AToM package. Raw characteristic angles are depicted in panel~(a). Tracked modes are depicted in panel~(b). Due to the complexity of original data ($100$~modes), only one mode from each degenerated eigenspace is depicted. This explains the missing data in panel~(a), highlighted by the shaded ovals. Notice the depicted difference in panel~(b) for connected and disconnected modes.}
		\label{fig8:numMFatom}
	\end{center}
\end{figure}

The results from the AToM package are depicted in Fig.~\ref{fig8:numMFatom}. The panel~(a) shows the raw data acquired from the Implicitly Restarted Arnoldi method \cite{Saad_NumericalMethodsForLargeEigenvalueProblems} (\texttt{eigs} in Matlab). This method was advantageously utilized since only the first $100$~modes are required at each frequency. The gray-colored ellipses highlight two exemplary samples at which the data are missing. The panel~(b) shows the modal data after the tracking procedure. A careful inspection reveals a couple of disconnected modes, one incomplete mode (missing data are highlighted by the gray-colored ellipse) and one missing inductive mode (depicted by the red dashed line). The systematic frequency shift between analytically predicted and numerically calculated internal resonances can be attributed to finite meshing is a consequence of the slightly smaller electrical size of the mesh grid.

\begin{figure}[t]
	\begin{center}
		\includegraphics[width=\figwidth cm]{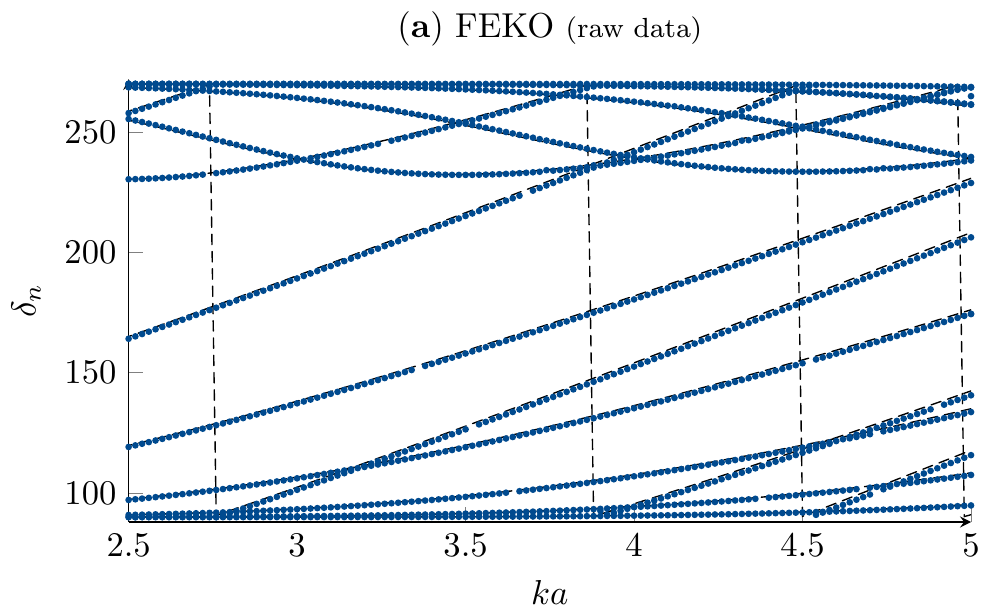}
		
		\vspace{0.3cm}
		\includegraphics[width=\figwidth cm]{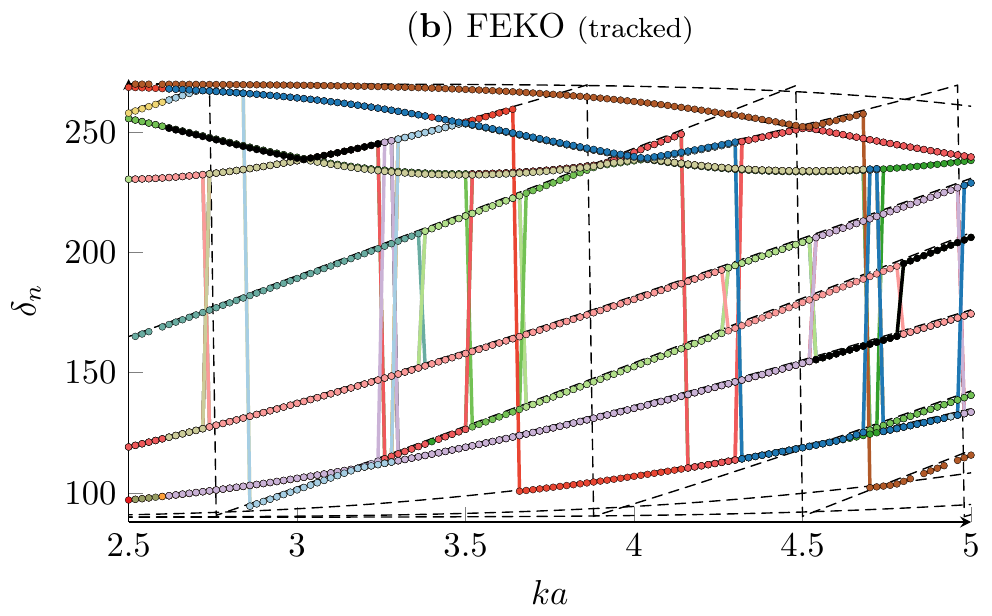}
		\caption{Comparison of raw (untracked) and tracked characteristic angles for the FEKO package.}
		\label{fig9:numMFFEKO}
	\end{center}
\end{figure}

The results from the FEKO package are depicted in Fig.~\ref{fig9:numMFFEKO}. The raw data in panel~(a) are similar to those found with the AToM package, however, the tracked modes in panel~(b) are far from perfect.

\begin{figure}[t]
	\begin{center}
		\includegraphics[width=\figwidth cm]{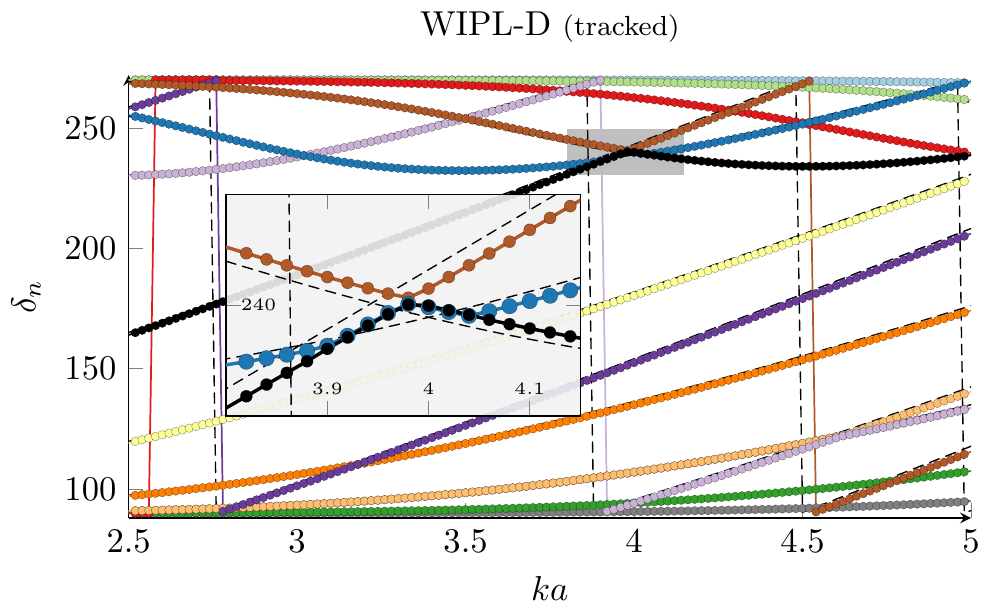}
		\caption{Tracked characteristic angles for the WIPL-D package. The region highlighted by gray color is enlarged in the inset and shows that selected modes are not tracked perfectly.}
		\label{fig9:numMmWIPLD}
	\end{center}
\end{figure}

The next analyzed package is WIPL-D which uses a quadrilateral mesh grid with higher-order basis functions which is why its results cannot be directly compared with other packages. However, the results seem promising, see Fig.~\ref{fig9:numMmWIPLD}, except at those places where the crossing-avoidances were incorrectly detected (see the inset). 

\begin{figure}[t]
	\begin{center}
		\includegraphics[width=\figwidth cm]{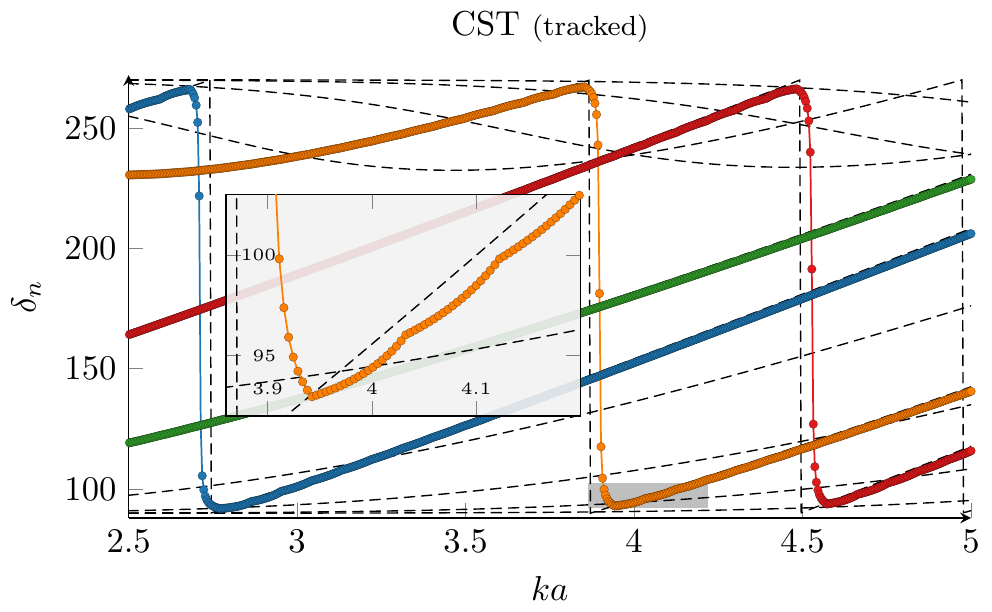}
		\caption{Tracked characteristic angles for the CST package. The region highlighted by gray color is enlarged in the inset and shows that data are highly interpolated.}
		\label{fig10:numMFCST}
	\end{center}
\end{figure}

The last package to undergo testing was CST-MWS. In comparison to the previous packages, CST-MWS uses CFIE for CMs analysis. Only the first four modes were found and, even though $226$~frequency samples were required, $1000$~interpolated values were returned, see Fig.~\ref{fig10:numMFCST}. The effect of the interpolation is evident from the enlarged section. The tracking procedure seems to be relatively computationally demanding as only $15$~modes could be calculated and only four modes were found. On the other hand, these modes are well-tracked.

\subsection{Test~\#3}
\label{Results3}
\begin{figure}[t]
	\begin{center}
		\includegraphics[width=\figwidth cm]{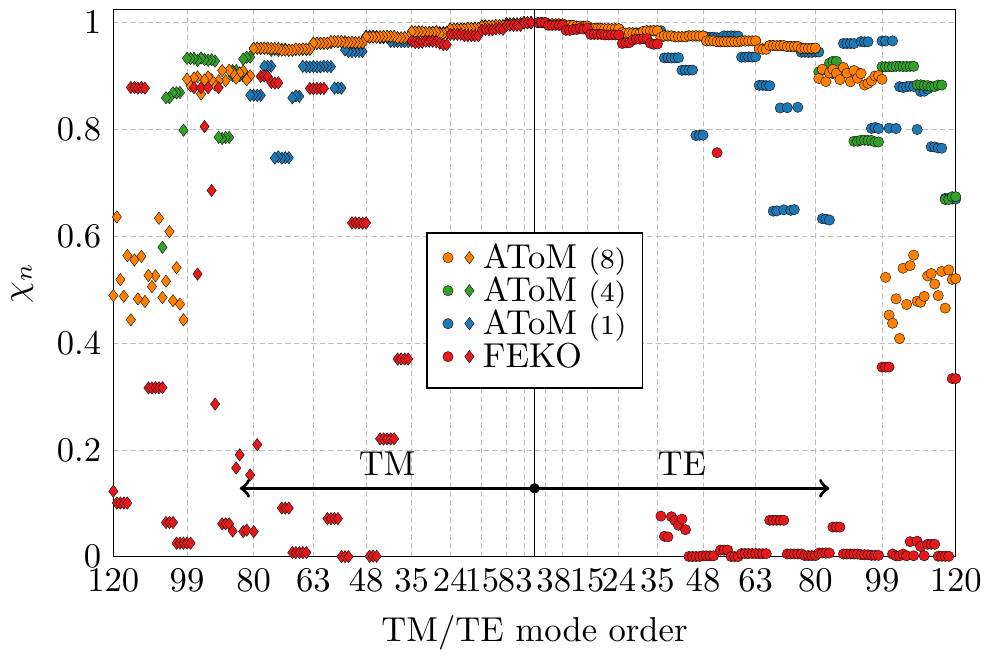}
		\caption{Similarity of numerically evaluated characteristic currents for a spherical shell discretized into $500$~triangles and an analytically known current from~\eqref{eq:JTE} and~\eqref{eq:JTM}. The electrical size is~\mbox{$ka = 1.5$}. Coefficients $\chi_n$ were calculated according to~\eqref{eq:JJtestN} for three different orders of the Gaussian quadrature rule~$\left\{1,4,8\right\}$ used in the AToM package to calculate impedance matrices and for the FEKO package.}
		\label{fig11:JJcorr750}
	\end{center}
\end{figure}

\begin{figure}[t]
	\begin{center}
		\includegraphics[width=\figwidth cm]{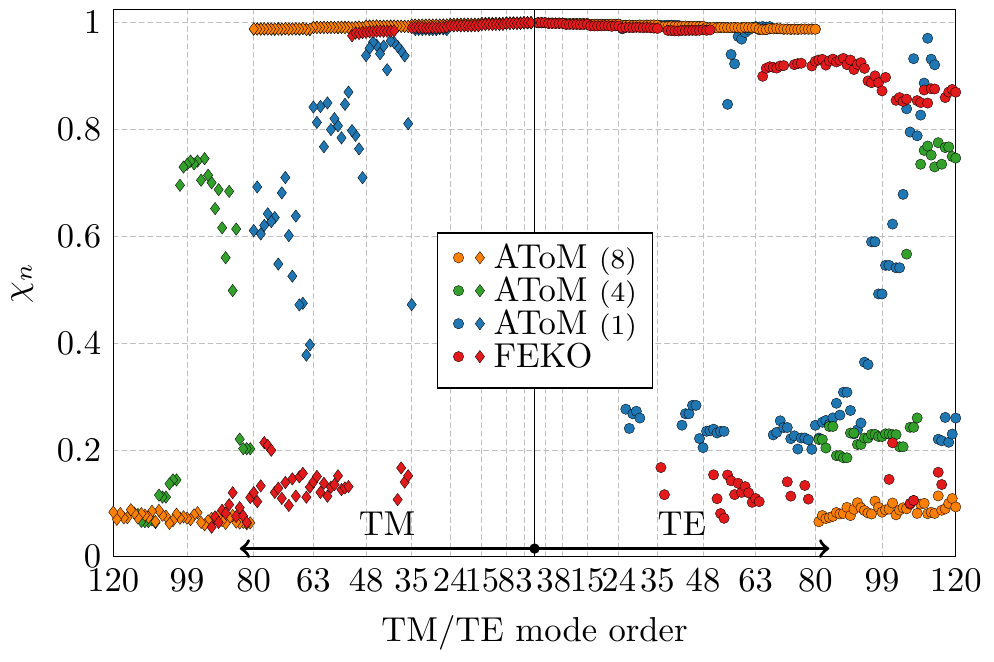}
		\caption{Same similarity study as in Fig.~\ref{fig11:JJcorr750} which was for a spherical shell discretized into $2200$~triangles.}
		\label{fig11:JJcorr3300}
	\end{center}
\end{figure}

The third test was performed only for the AToM and FEKO packages at~\mbox{$ka=1.5$} for a mesh grid composed of~$500$ and~$2200$ triangles, see Fig.~\ref{fig11:JJcorr750} and Fig.~\ref{fig11:JJcorr3300}. The similarity coefficient~$\chi_n$ from \eqref{eq:JJtestN} decreased significantly faster for the poorer mesh grid, however, the number of sufficiently represented modes, say those with \mbox{$\chi_n > 0.9$}, was similar for both mesh grids. The influence of the higher-order quadrature rule is obvious and the number of well-defined currents, approximately the first $80$~TM and TE modes, including degeneracies, corresponds perfectly with the number of precisely calculated eigenvalues, see Fig.~\ref{fig7:numSFLam05}.

\subsection{Test~\#4}
\label{Results4}

\begin{figure}[t]
	\begin{center}
		\includegraphics[width=\figwidth cm]{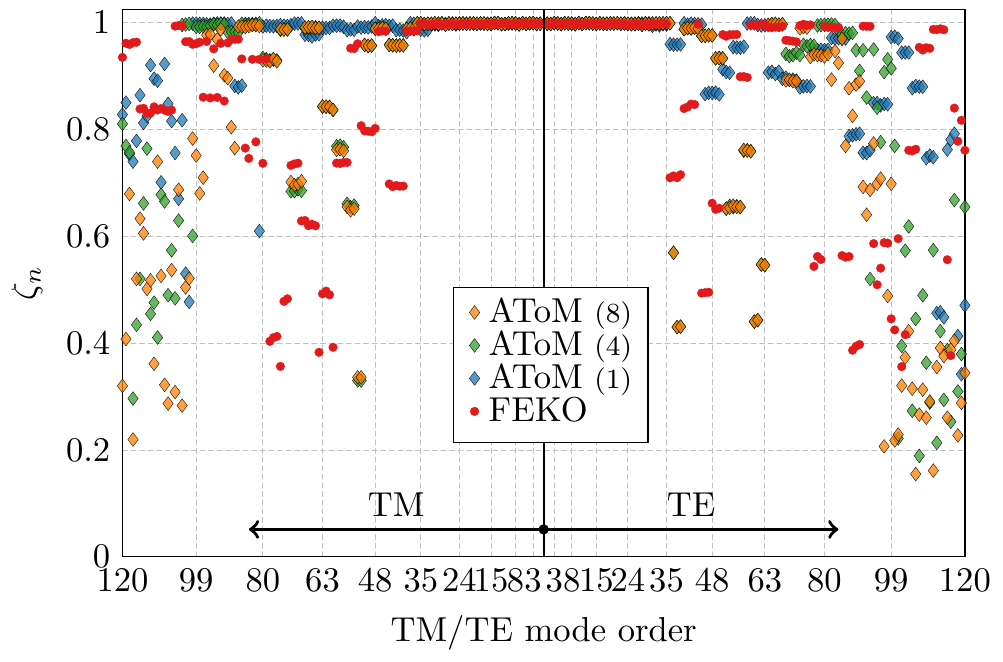}
		\caption{Similarity of numerically and analytically evaluated characteristic far-fields for a spherical shell discretized into $500$~triangles at electrical size~\mbox{$ka = 1.5$}. Coefficients $\zeta_n$ were calculated according to~\eqref{eq:test3eq6}.}
		\label{fig16:FF750edg1}
	\end{center}
\end{figure}

\begin{figure}[t]
	\begin{center}
		\includegraphics[width=\figwidth cm]{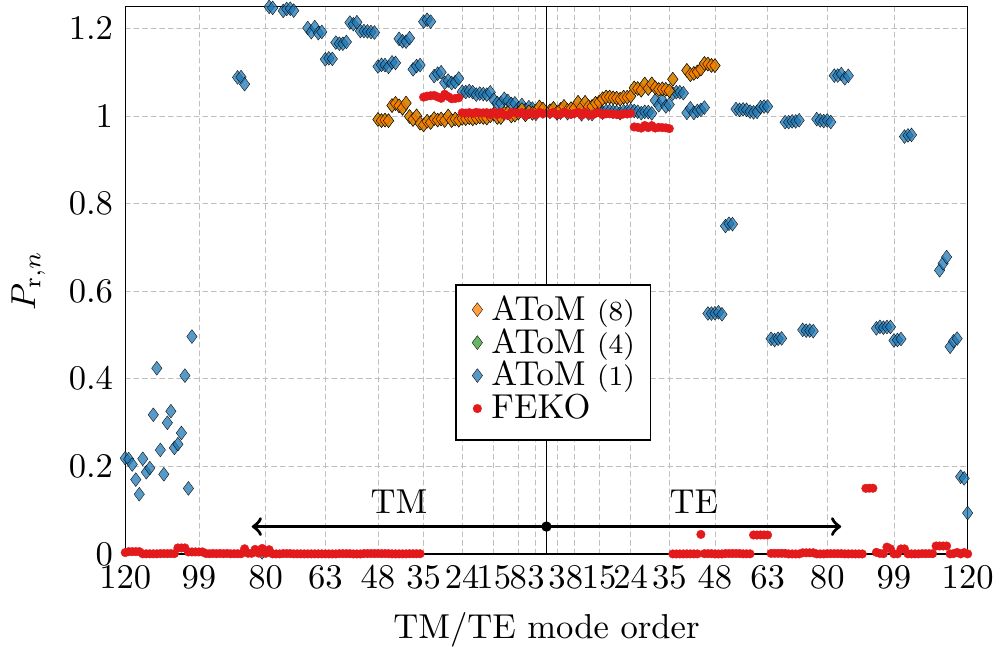}
		\caption{Far-field radiated power (denominator of \eqref{eq:test3eq6}) of $n$-th characteristic mode. The same input data as for Fig.~\ref{fig16:FF750edg1} have been used. Note that the radiated power of some higher order modes is out of the depicted range.}
		\label{fig17:FF750edg2}
	\end{center}
\end{figure}

Test metric $\zeta_n$ from \eqref{eq:test3eq6} was evaluated at electrical size $ka=1.5$ for data obtained from the AToM and FEKO packages, with results depicted in Fig.~\ref{fig16:FF750edg1} and Fig.~\ref{fig17:FF750edg2}. The test was not performed using the CST-MWS package since it internally normalizes characteristic modes via~$L^2$ norm and not via~\eqref{eq:CMdef}. The results could, therefore, not be compared with others. The CEM One package does not have characteristic far-fields implemented.

For proper assessment, the results in both figures should be compared simultaneously since only simultaneous compliance with~\mbox{$\zeta_n \approx 1$} and~\mbox{$P_{\mathrm{r},n} \approx 1$} is sufficient to pass the test, with~\mbox{$P_{\mathrm{r},n}$} being the far-field radiated power of the $n$-th characteristic mode, \ie{}, the denominator of \eqref{eq:test3eq6}. The normalization of radiated power may seem to be automatic, but for higher order modes the normalization \eqref{eq:CMdef}, performed numerically via~\mbox{$\Ivec^\herm \M{R} \Ivec = 2$}, does not imply unitary radiated power in far-field.

Only modes with real eigenvalues and positive radiated power were taken in the case of AToM. Data from FEKO were left in their original form. In all cases, the integration~\eqref{eq:test3eq4} was discretized in \mbox{$N_\vartheta = N_\varphi = 100$}~points. This angular discretization was checked by artificially testing analytical modes which resulted in \mbox{$\zeta_n \approx 1$} and \mbox{$P_{\mathrm{r},n} \approx 1$} for all analytical modes within the depicted range.

The performance in all tested cases is quite unsatisfactory, but generally corresponds to the other tests. The results suggest that far-field decomposition into more than five TE modes and five TM modes (not including degenerations) of a spherical shell is unsafe within the used triangularization of a sphere and double precision. Higher quadrature rules and careful singularity treatment, seen in the case of AToM with an $8$-th order of quadrature, add more proper modes, but the gain is not as high as in other tests.

\section{Conclusion}
\label{Sec7}

A set of sensitive benchmarks applicable to the majority of contemporary method of moments packages was proposed. The major advantages are formal simplicity, the low number of necessary inputs, the independence on particular feeding model and minimal required post-processing.

All tested packages showed satisfactory performance in calculating characteristic eigenvalues at a single frequency, although some of them were limited out of the single numerical precision range. Without exception, however, they performed well below theoretically achievable results. In the case of modal tracking, the differences were more severe and some packages showed unsatisfactory results. It has been demonstrated that factors, such as singularity treatment, high-order quadrature rules and used floating-point numerical precision, greatly influence the results. A persisting problem is also the numerical ill-posedness of the real part of impedance matrix caused by dominance of only couple of modes. The application of any technique increasing its dynamical range while preserving its precision is, therefore, of interest since it may significantly improve the number of correctly found characteristic modes. 

Future work should be aimed toward analytical characteristic modes of more complex shapes which could further stress the precision of available method of moments codes. Promising candidates are surfaces which can be described in conical or spheroidal coordinate systems.

\appendices

\section{FEKO Setup}
\label{appFEKO}

FEKO (ver.~14.0-273612,~\cite{feko}) was used with the following settings: a mesh structure was imported into the software using the Nastran file format \cite{nastran} and the CMs and far-fields were chosen as a request for the FEKO solver. Data from FEKO were acquired using *.out, *.os, *.mat and *.ffe files. The impedance matrices were imported by an in-house wrapper \cite{IDA}. Double precision was enabled for data storage in solver settings. The number of modes used for each test were as follows: Test~\#2 $99$~modes, Test~\#3 $500$~modes and Test~\#4 $300$~modes, respectively.

\section{WIPL-D Setup}
\label{appWIPLD}
WIPL-D (v13,~\cite{wipld}) uses higher-order basis functions with a quadrilateral mesh grid. Therefore, individual meshing was used. Surface angle tolerance was set to $15$~degrees. Additionally, a non-symmetrical mesh was created in order to obtain better stability of mode calculations. An integral solver with CM decomposition was utilized with double precision and enhanced-1 for Integral Accuracy. The matrices for Test~\#1 were delivered by the developer of the WIPL-D package and data for Test~\#2 included first $100$~modes.

\section{CST-MWS Setup}
\label{appCSTMWS}

CST-MWS (ver.~2016.7,~\cite{cst}) setting was as follows: the spherical shell was created using a sphere modeling tool and converted into a sheet. The parameter cells per wavelength was set to $5.25$ in general mesh properties which led to a mesh grid of~$500$~triangular elements. The Integral equation solver was chosen and the number of requested modes was set to 16. It is worth noting that regardless of the choice of number of frequency samples in CMA accuracy settings, CST-MWS always used~$1001$~frequency samples.

\section{CEM One Setup}
\label{appCEMOne}

CEM One (2015.2, \cite{CemOne}) setting was as follows: a mesh grid was imported using the Nastran file format \cite{nastran}. The impedance matrix was obtained by the \Quot{Save System Matrix} in the Final Output Parameters. Quadrature order was set in a *.dat file as:
\begin{verbatim}
#RUMSEY
simple 3 6
double 3 6
\end{verbatim}
The first number represents the quadrature order for far elements and the second number represents the quadrature order for singularities. The impedance matrix was saved using a \texttt{ncdump} command in the E-Field command Prompt. 

\section{AToM Setup}
\label{appAToM}

AToM (pre-product ver.) setting was as follows: a mesh grid was imported using the Nastran file format \cite{nastran}. The number of modes was set manually to $100$~for Test~\#2 (the \texttt{eigs} routine was used) and kept as the maximum for the other tests (the \texttt{eig} routine). In AToM, RWG basis functions with the Galerkin procedure are adapted~\cite{RaoWiltonGlisson_ElectromagneticScatteringBySurfacesOfArbitraryShape}. The Gaussian quadrature is implemented according to~\cite{Dunavant_HighDegreeEfficientGQR} and singularity treatment is implemented from~\cite{EibertHansen_OnTheCalculationOfPotenticalIntegralsForLinearSourceDistributionsOnTriangularDomains}. The tracking of the CMs follows~\cite{CapekHazdraHamouzEichler_AMethodForTrackingCharNumbersAndVectors}, including some recently added adaptive strategies~\cite{SafinManteuffel_AdvancedEigenvalueTrackingofCM}. Since AToM is written in Matlab, no other import procedures were needed.

\section*{Acknowledgement}
The authors would like to thank Vladimir Sedenka from BUT and Michal Masek from CTU in Prague for implementing the method of moments and characteristic modes solvers within the AToM package, and would also like to thank to Gerhard Kristensson from Lund University for a fruitful discussion concerning separable systems, Kurt Schab from NC State University, Doruk Tayli from Lund University and Jasmin Music from \mbox{WIPL-D} for providing the impedance matrices.

\bibliographystyle{IEEEtran}
\bibliography{references_LIST_UpToDate}

\begin{IEEEbiography}[{\includegraphics[width=1in,height=1.25in,clip,keepaspectratio]{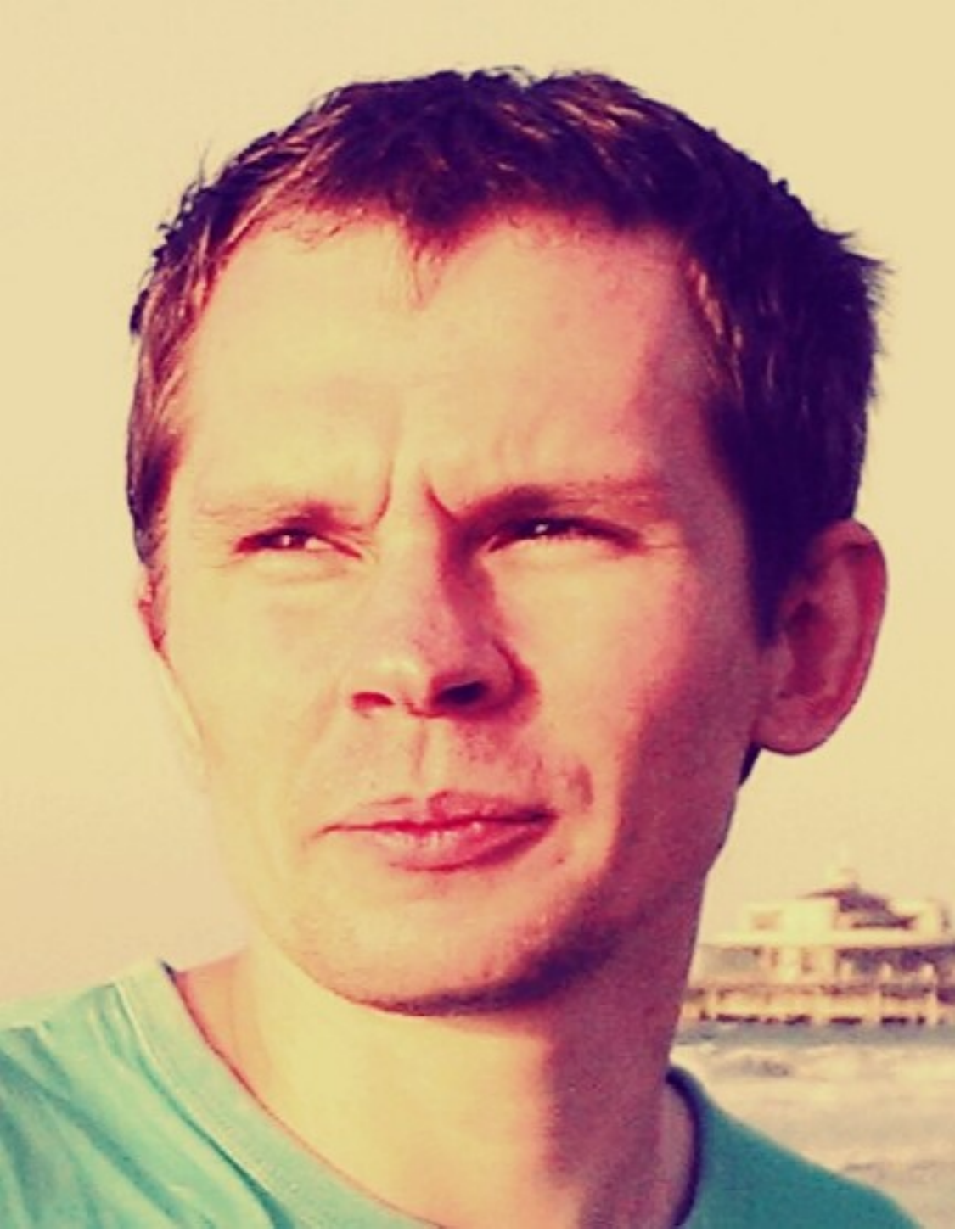}}]{Miloslav Capek}(S'09, M'14) received his M.Sc. degree in Electrical Engineering from the Czech Technical University, Czech Republic, in 2009, and his Ph.D. degree from the same University, in 2014. Currently, he is a researcher with the Department of Electromagnetic Field, CTU-FEE.
	
He leads the development of the AToM (Antenna Toolbox for Matlab) package. His research interests are in the area of electromagnetic theory, electrically small antennas, numerical techniques, fractal geometry and optimization. He authored or co-authored over 50 journal and conference papers.
	
Dr. Capek is member of Radioengineering Society, regional delegate of EurAAP, and Associate Editor of Radioengineering.
\end{IEEEbiography}

\begin{IEEEbiography}[{\includegraphics[width=1in,height=1.25in,clip,keepaspectratio]{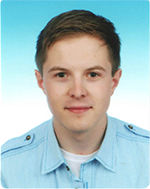}}]{Vit Losenicky}
received the M.Sc. degree in electrical engineering from the Czech Technical University in Prague, Czech Republic, in 2016. He is now working towards his Ph.D. degree in the area of electrically small antennas.
\end{IEEEbiography}

\begin{IEEEbiography}[{\includegraphics[width=1in,height=1.25in,clip,keepaspectratio]{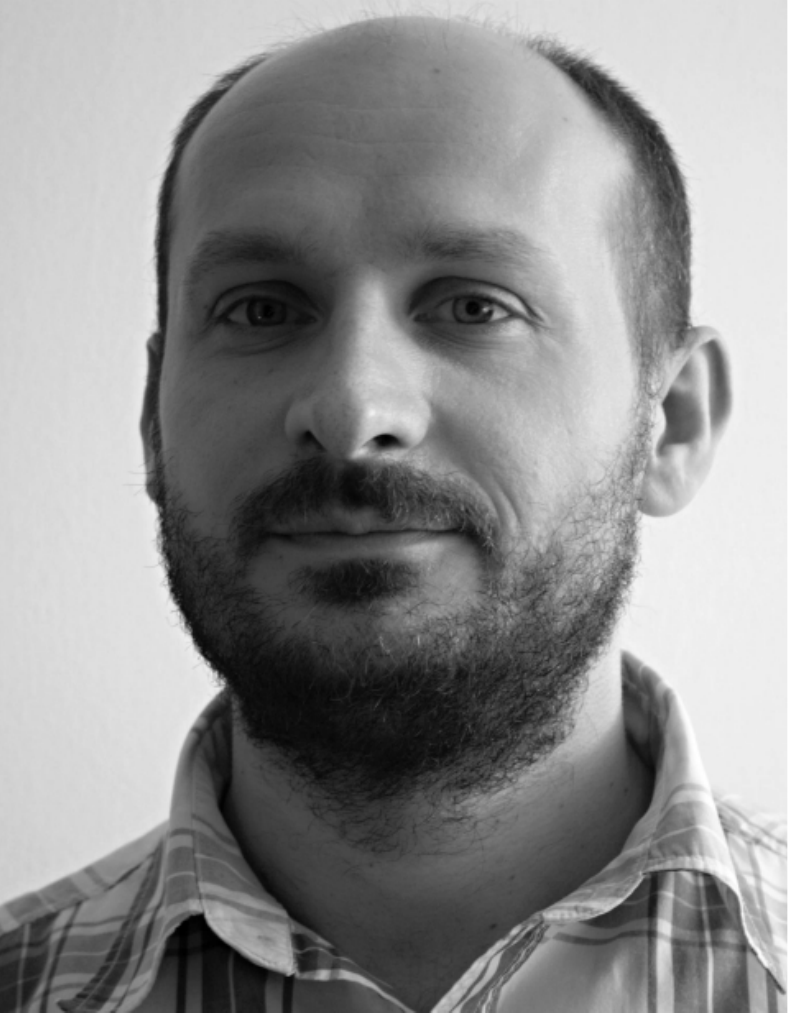}}]{Lukas Jelinek}
received his Ph.D. degree from the Czech Technical University in Prague, Czech Republic, in 2006. In 2015 he was appointed Associate Professor at the Department of Electromagnetic Field at the same university.

His research interests include wave propagation in complex media, general field theory, numerical techniques and optimization.
\end{IEEEbiography}

\begin{IEEEbiography}[{\includegraphics[width=1in,height=1.25in,clip,keepaspectratio]{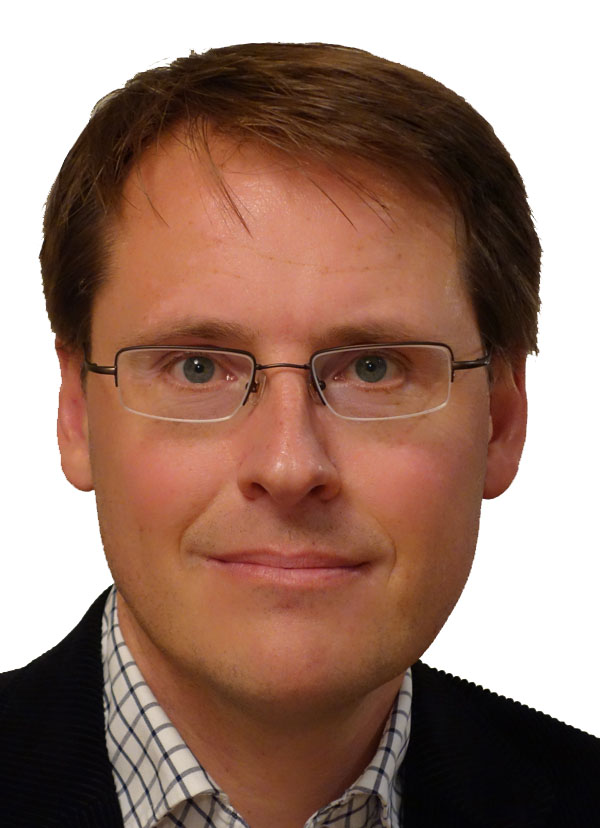}}]{Mats Gustafsson}
received the M.Sc. degree in Engineering Physics 1994, the Ph.D. degree in Electromagnetic Theory 2000, was appointed Docent 2005, and Professor of Electromagnetic Theory 2011, all from Lund University, Sweden. 
	
He co-founded the company Phase holographic imaging AB in 2004. His research interests are in scattering and antenna theory and inverse scattering and imaging. He has written over 80 peer reviewed journal papers and over 100 conference papers. Prof. Gustafsson received the IEEE Schelkunoff Transactions Prize Paper Award 2010 and Best Paper Awards at EuCAP 2007 and 2013. He served as an IEEE AP-S Distinguished Lecturer for 2013-15.
\end{IEEEbiography}
\end{document}